\documentstyle[epsfig]{mn}

\newcommand\lta{\mathrel{\rlap{\lower 3pt\hbox{$\mathchar"218$}}
     \raise 2.0pt\hbox{$\mathchar"13C$}}}
\newcommand\gta{\mathrel{\rlap{\lower 3pt\hbox{$\mathchar"218$}}
     \raise 2.0pt\hbox{$\mathchar"13E$}}}
\newcommand\kms{km~s$^{-1}$}

\newcommand\etal{{et~al.}} 
\newcommand\sigth{\ifmmode \sigma_{\rm th}\else$\sigma_{\rm th}$\fi}
\newcommand\sigv{\ifmmode \sigma_v\else$\sigma_v$\fi}
\newcommand\mM{\ifmmode(m{-}M)\else$(m{-}M)$\fi}
\newcommand\msun{\ifmmode{\hbox{M$_\odot$}}\else{M$_\odot$}\fi}

\newcommand\bi{\ifmmode \beta_{I}\else$\beta_I$\fi}
\newcommand\bo{\ifmmode \beta_{O}\else$\beta_O$\fi}
\newcommand\mgii{\ifmmode\hbox{Mg}_2\else{Mg$_2$}\fi}

\def\PD{\hbox{\sc pd}}
\def\hkpc{$h^{-1}\,$kpc}

\def\mi{\ifmmode\overline{m}_I\else$\overline{m}_I$\fi}
\def\mv{\ifmmode\overline{m}_V\else$\overline{m}_V$\fi}
\def\mbar{\ifmmode\overline{m}\else$\overline{m}$\fi}
\def\Mbar{\ifmmode\overline{M}\else$\overline{M}$\fi}
\def\lbar{\ifmmode\overline{L}\else$\overline{L}$\fi}
\def\ibar{\ifmmode\overline{I}\else$\overline{I}$\fi}
\def\vbar{\ifmmode\overline{V}\else$\overline{V}$\fi}
\def\vbib{\ifmmode(\overline{V}{-}\overline{I})\else$(\overline{V}{-}\overline{I})$\fi}
\def\vbkb{\ifmmode(\overline{V}{-}\overline{K})\else$(\overline{V}{-}\overline{K})$\fi}
\def\ibkb{\ifmmode(\overline{I}{-}\overline{K})\else$(\overline{I}{-}\overline{K})$\fi}
\def\Mi{\ifmmode\overline{M}_I\else$\overline{M}_I$\fi}
\def\Miz{\ifmmode\overline{M}_I^0\else$\overline{M}_I^0$\fi}
\def\Mv{\ifmmode\overline{M}_V\else$\overline{M}_V$\fi}
\def\vi{\ifmmode(V{-}I)\else$(V{-}I)$\fi}
\def\viz{\ifmmode(V{-}I)_0\else$(V{-}I)_0$\fi}
\def\bv{\ifmmode(B{-}V)\else$(B{-}V)$\fi}
\def\bvz{\ifmmode(B{-}V)_0\else$(B{-}V)_0$\fi}
\def\vimg{$(V{-}I)_0$--{\rm Mg}$_2$}
\def\dn{\ifmmode D_n\hbox{-}\sigma\else$D_n\hbox{-}\sigma$\fi}
\def\Nbar{\ifmmode\overline{N}\else$\overline{N}$\fi}
\def\ho{\ifmmode H_0\else$H_0$\fi}
\def\lsig{\ifmmode \log(\sigma)\else$\log(\sigma)$\fi}
\def\xfp{\ifmmode X_{\rm FP}\else$X_{\rm FP}$\fi}
\def\sbe{\ifmmode \langle\mu\rangle_e \else $\langle\mu\rangle_e$\fi}
\def\sbeR{\ifmmode \langle\mu_R\rangle_e \else $\langle\mu_R\rangle_e$\fi}
\def\aj{AJ}
\def\apj{ApJ}
\def\apjl{ApJ}
\def\apjs{ApJS}
\def\mnras{MNRAS}

\def\pasp{PASP}
\def\aap{A\&A}

\title[Synthesis of FP and SBF]{A Synthesis of Data from Fundamental Plane and Surface Brightness Fluctuation Surveys}

\author[J.\ P.\ Blakeslee et al.]{John P. Blakeslee,$^{1,2}$
John R.\ Lucey,$^{2}$ Brian J.\ Barris,$^{3}$
Michael J.\ Hudson,$^{4}$ \newauthor and John L.\ Tonry$^{3}$ \\
$^1${Department of Physics and Astronomy,
Johns Hopkins University, Baltimore, MD 21218, U.S.A.; jpb@pha.jhu.edu}\\
$^2${Department of Physics, University of Durham, South Road, 
Durham, DH1\,3LE, United Kingdom; John.Lucey@durham.ac.uk}\\
$^3${Institute for Astronomy, University of Hawaii,
2680 Woodlawn Drive, Honolulu, HI 96822, U.S.A.; barris,jt@ifa.hawaii.edu}\\
$^4${Department of Physics, 
University of Waterloo, ON, N2L\,3G1, Canada; mjhudson@uwaterloo.ca}\\
}

\date{Submitted\,--- 30 March 2001.~  Accepted\,--- 5 July 2001.}
\pagerange{\pageref{firstpage}--\pageref{lastpage}}\pubyear{2001}

\begin{document}

\maketitle \label{firstpage}

\begin{abstract}
We perform a series of comparisons between distance-independent
photometric and spectroscopic properties used in the surface brightness
fluctuations (SBF) and fundamental plane (FP) methods of early-type galaxy
distance estimation. The data are taken from two recent surveys: the SBF
Survey of Galaxy Distances and the Streaming Motions of Abell Clusters
(SMAC) FP~survey.
We derive a relation between \viz\ colour and \mgii\ index using nearly
200 galaxies and discuss implications for Galactic extinction estimates
and early-type galaxy stellar populations.
We find that the reddenings from Schlegel \etal\ (1998) for galaxies with
$E(B{-}V) \gta 0.2$ mag appear to be overestimated by 5--10\%, but we do
not find significant evidence for large-scale dipole errors in the extinction map.
In comparison to stellar population models having solar elemental
abundance ratios, the galaxies in our sample are generally too blue at a
given \mgii; we ascribe this to the well-known enhancement of the
$\alpha$-elements in luminous early-type galaxies.
We confirm a tight relation between stellar velocity dispersion~$\sigma$
and the SBF `fluctuation count' parameter \Nbar, which is a
luminosity-weighted measure of the total number of stars in a galaxy.
The correlation between \Nbar\ and $\sigma$ is even tighter than that
between \mgii\ and~$\sigma$.
Finally, we derive FP photometric parameters for
280 galaxies from the SBF survey data set. Comparisons with
external sources allow us to estimate the errors on these parameters
and derive the correction necessary to bring them onto the SMAC system.
The data are used in a companion paper 
which compares the distances derived from the FP and SBF methods.
\end{abstract}
\begin{keywords}
galaxies: distances and redshifts ---
galaxies: elliptical and lenticular, cD  ---
galaxies: fundamental parameters ---
galaxies: stellar content
\end{keywords}

\section{Introduction}

However violent or disturbed in youth, mature elliptical galaxies
are remarkably well-behaved members of the celestial pageant.
Their photometric and structural properties obey well-defined relations
over a very large range in mass and luminosity.  The
correlations of distance-dependent properties with distance-independent
ones then allows for the derivation of galaxy distances.

The fundamental plane (FP) (Dressler \etal\ 1987; Djorgovski \& Davis
1987) works as a distance indicator because elliptical galaxies are
dynamically hot systems obeying the virial theorem and have fairly
similar, mainly old, stellar populations.
The distance-dependent quantity is a combination of the galaxy effective
radius and surface brightness (or just the galaxy luminosity for the
Faber-Jackson [1976] relation), and the distance-independent one is the stellar
velocity dispersion $\sigma$.  Sometimes a stellar population term,
usually the \mgii\ index, is also incorporated into the FP (e.g., Guzm\'an
\& Lucey 1993). 

The surface brightness fluctuations (SBF) distance method (Tonry \&
Schneider 1988) relies solely on the stellar population properties of old
stellar systems; this enables Cepheid calibration via SBF measurements in
spiral bulges.  The distance-dependent quantity for SBF is the
luminosity-weighted mean luminosity of the stellar population and the
distance-independent one is the \vi\ colour, or possibly some other stellar
population indicator.

This paper examines and compares the various photometric and spectroscopic
parameters used in the SBF and FP distance methods.  We first discuss the
data samples used for this study before exploring correlations between the
distance-independent parameters which go into the methods. Special
attention is given to potential systematic effects that could bias the
distance estimates.  We then describe our procedure for deriving the
distance-dependent FP photometric parameters from SBF survey imaging data
and compare the results with external data in order to evaluate their
accuracy.  We include a data table listing both the FP and SBF parameters;
these are used in a companion paper (Blakeslee \etal\ 2002,
hereafter Paper~II) which makes detailed
comparisons of the distance estimates for individual galaxies from the two
methods.

\section{Distance-Independent Comparisons}

For this study, we combine data from two recent early-type galaxy distance
surveys: the SBF Survey of Galaxy Distances and the Streaming Motions of
Abell Clusters (SMAC) project.  The SBF survey data and calibration are
described in detail by Tonry \etal\ (1997, 2000, hereafter SBF-I and
SBF-II), with further analysis of the data given by Blakeslee \etal\
(1999, SBF-III).  SBF survey data for 300 galaxies within $c{z}\lta4000$
\kms\ are tabulated by Tonry \etal\ (2001, SBF-IV).  The SMAC project is a
cluster FP survey which combines new and literature data into a common
system.  As well as providing standardised parameters for cluster
galaxies, the SMAC project derived improved values for many nearby
non-cluster galaxies.  The new photometric and spectroscopic data are
presented by Smith \etal\ (2000, 2001, hereafter SMAC-I and SMAC-II),
and the full SMAC data set is  tabulated by Hudson \etal\ (2001,
SMAC-III).  A preliminary analysis of the peculiar velocity data was given by
Hudson \etal\ (1999).  Here, we match data for galaxies
appearing in both surveys and discuss correlations between the various
distance-independent galaxy properties.

\subsection{The \vimg--$\sigma$ Relations\label{ssec:massmetal}}
We matched all galaxies with \vi\ colours from the SBF survey (SBF-IV)
against the SMAC spectroscopic catalogue of homogenized \mgii\ and
$\sigma$ measurements.  The spectroscopic measurements have been
normalized to a standard aperture of radius $r_{\rm ap} = 0.6$ \hkpc\ by
assuming a fixed radial gradient from Jorgensen \etal\ (1995) for all the
galaxies and using the observed velocities to approximate distances (see
SMAC-III).  The \vi\ colours, however, are derived from the regions within
the galaxies that were used for the SBF analysis.  Because the galaxy
centres are usually saturated in the SBF observations and the central
regions are frequently affected by dust, these colours roughly correspond
to the colour near one effective radius $r\approx R_e$ (aperture effects
on the \vimg\ relation are discussed in \S\ref{ssec:heredity}).
Galactic extinction values and ratios from Schlegel, Finkbeiner, \& Davis
(1998, hereafter SFD) were used in deriving the reddening-corrected 
\viz\ colours (see SBF-II).  The adopted reddening law gives
$E(V{-}I) = 1.278E(B{-}V)$.

A total of 209 galaxies were
found to have both \vi\ and \mgii\ data. One of these is M32, which
has an extraordinarily large estimated aperture correction,
while two others (NGC\,404 and NGC\,5102) have
$\mgii\lta0.0$. We omit these three galaxies, leaving a sample 
of 206, all but two (E358-059 and NGC\,4468) of which also have 
SMAC values for $\sigma$.   The mean observational uncertainties
are 0.018\,mag in \vi\ (not including reddening errors) and
0.007\,mag in \mgii; SBF-I and SMAC-III describe how the errors
were estimated from comparisons of repeat observations.

\begin{figure}
\medskip\vbox{\centering\leavevmode\hbox{
\epsfxsize=8.0cm\epsffile{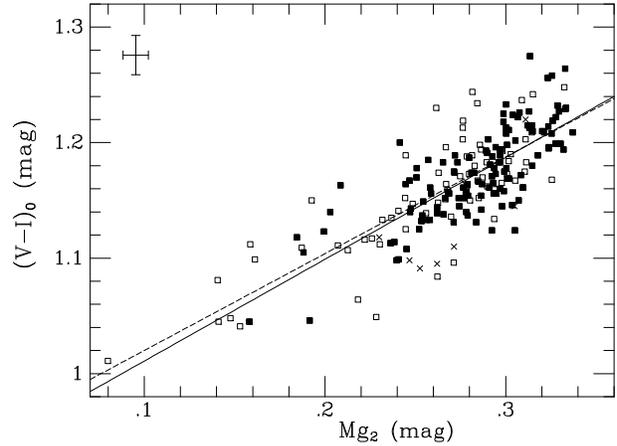} }\caption{\small \viz\ colour
from the SBF survey is plotted against \mgii\ index from the SMAC survey
for galaxies in common between the two data sets.  Filled squares
represent elliptical galaxies ($T$-type = $-5$), open squares are S0
galaxies ($T > -5$), and the crosses represent galaxies with relatively
high Galactic extinctions ($A_V > 0.5$~mag from SFD). Median 
measurement errors are shown at upper left.  The heavily
reddened galaxies are too blue at a given Mg$_2$, indicating that
their reddenings have been overestimated. The solid and dashed lines
are fits to the elliptical and elliptical+S0 samples, respectively.
\label{fig:mg2vimain}}}
\end{figure}

\begin{figure}
\medskip\vbox{\centering\leavevmode\hbox{
\epsfxsize=8.6cm\epsffile{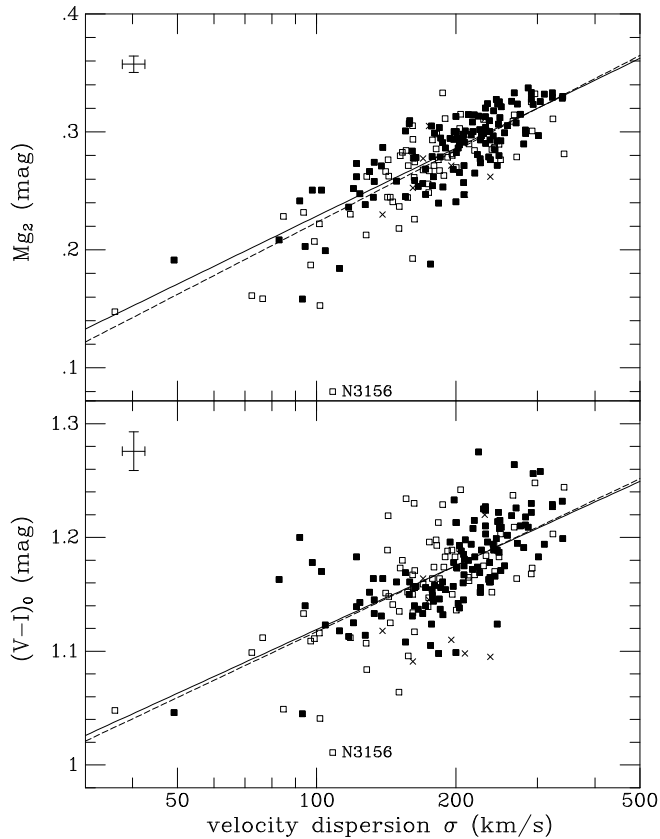} }\caption{\small Mg$_2$ index
from the SMAC survey and \viz\ colour from the SBF survey are plotted
against the central stellar velocity dispersion $\sigma$ from SMAC.
Symbols and lines are as in Figure~\ref{fig:mg2vimain}, with
median error bars shown at upper left in each panel.
The S0 galaxy NGC\,3156 with \mgii$\,=\,$0.08 is the biggest outlier
and has been omitted from the fits.
Galaxies with $A_V > 0.5$~mag (crosses) are too blue
in \viz\ with respect to the average relation, again indicating
that their reddenings are overestimated.  The $\mgii-\sigma$
relation is not sensitive to reddening errors.
\label{fig:mgisig}}}
\end{figure}

Figure~\ref{fig:mg2vimain} shows the correlation of \viz\ with \mgii\
for the cross-matched data set.  We find that ellipticals
(filled squares, defined as having morphological type $T{\,=\,}{-}5$ in the RC3
[de\,Vaucouleurs et al.\ 1991]), and S0s (open squares,
effectively $T{\;>\;}{-}5$ for this sample) obey the same \vimg\ relation.  
This is expected for a pair of purely stellar population parameters with 
similar age--metallicity degeneracy properties (e.g., Worthey 1994).
However, the 8 highest extinction galaxies (crosses in Figure~\ref{fig:mg2vimain}),
which have $E(V{-}I)>0.20$ and $A_V>0.52$,
preferentially lie below the mean \vimg\ relation.
The least-squares fits for galaxies with
$E(V{-}I)<0.20$ and $\mgii>0.1$ are given by
\begin{eqnarray}
&(V{-}I)_0\;=\;(0.935\pm0.014) + (0.843\pm0.049)\,\mgii\,,&\label{eq:vimg_all}\\
&(V{-}I)_0\;=\;(0.923\pm0.019) + (0.881\pm0.067)\,\mgii\,,&\label{eq:vimg_ell} 
\end{eqnarray}
for all morphological types (197 galaxies, rms scatter 0.0275 mag) and
for just the ellipticals (125 galaxies, rms scatter 0.0257 mag),
respectively.  These fits are clearly identical within the errors.

The 72 S0s exhibit a marginally higher scatter of 0.030 mag, however
they also have larger \vi\ measurements uncertainties 
on average, 0.020 mag compared to 0.017 mag for the ellipticals.
Therefore, we assign no physical significance to the marginal
difference in scatter. If we perform a bivariate fit using the 
quoted uncertainties added in quadrature with a cosmic scatter term,
we find that the cosmic scatter in \viz\ at a fixed \mgii\ must
be 0.021\,mag to obtain a reduced $\chi^2$ of unity
(or, formally, 0.020 when just the ellipticals are considered
and 0.022 for just the S0s).
This cosmic scatter actually would include errors 
resulting from gradient and aperture effects, incorrect reddening
estimates, as well as true peculiarities of the stellar populations.
In the following section, we use the residuals from the \vimg\ relation 
to investigate problems with the extinction estimates.

Figure~\ref{fig:mgisig} plots \mgii\ and \viz\ against the stellar
velocity dispersion $\sigma$ for the cross-matched sample.  The
\hbox{\mgii--$\sigma$} relation has been discussed in detail by many
authors (e.g., Terlevich \etal\ 1981; Dressler \etal\ 1987; Guzm\'an
\etal\ 1992; Bender, Burstein, \& Faber 1993; Colless \etal\ 1999), and we
mainly include it here as a comparison to the $\viz$--$\sigma$ relation.
Excluding the discordant S0 NGC\,3156 which has $\mgii<0.1$, we find
\begin{equation}
\mgii\;=\;(-0.194\pm0.025) + (0.208\pm0.011)\,\log(\sigma)\,,
	\label{eq:mgsig_all}
\end{equation}
with an rms scatter of 0.022 mag for the full morphological sample
(0.021 mag for just the ellipticals), consistent with most previous studies.
For the \viz--$\sigma$ comparison,
we again find that galaxies with $E(V{-}I)>0.20$ mag appear
to have had their reddening corrections overestimated by SFD.  Excluding
these and NGC\,3156, we find
\begin{equation}
(V{-}I)_0\;=\;(0.733\pm0.033) + (0.192\pm0.015)\,\log(\sigma)\,,
	\label{eq:visig_all}
\end{equation}
with an rms scatter of 0.031 mag for the full morphological sample
(again, the elliptical subsample gives a near identical relation and a
scatter 0.030 mag).  We know of no other published derivations
of a \vi--$\log\sigma$ relation.  However,
Bender \etal\ (1993) presented a relation between
$(B{-}V)_0$ and $\log\sigma$, mainly using the highest quality
subsample of 7S data from Burstein \etal\ (1987).  
The slope of that relation is also near 0.2 and the scatter 
is 0.032 mag when the blue galaxies with $\bvz<0.8$
are omitted.   Bower, Lucey, \& Ellis (1992) examined the 
$(V{-}K)$--$\log\sigma$ relation and found scatters of 
0.032 and 0.050~mag for samples comprised of 17 Virgo 
and 31 Coma early-type galaxies, respectively.

\begin{figure}
\medskip\vbox{\centering\leavevmode\hbox{
\epsfxsize=8.0cm\epsffile{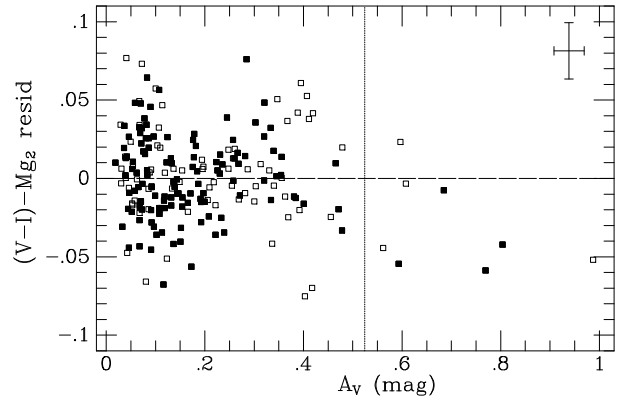} }\caption{\small Residuals from
a linear fit to the \vimg\ relation are plotted against the $V$-band
Galactic extinction value from SFD for ellipticals (filled squares) and
S0s (open squares).  Median error bars are shown at upper right, where the
$A_V$ error estimate is taken from SFD.
The vertical dotted line is drawn at $A_V = 0.52$
mag, or $E(V{-}I)=0.20$ mag; the eight galaxies at higher extinctions
(which were represented by crosses in Figures~\ref{fig:mg2vimain}
and~\ref{fig:mgisig}) tend to lie below the zero-residual line, indicating
that their extinctions have been overestimated.  
\label{fig:mg2viext}}}
\end{figure}

\subsection{Tests of the Galactic Extinction}
\label{ssec:galext}

We can use the \vimg\ relation as a test of
the Galactic extinction map.  Figure~\ref{fig:mg2viext} shows the
residuals with respect to Eq.\,(\ref{eq:vimg_ell}) and indicates
an apparent problem at the highest extinctions: 7 of 8 galaxies 
(and all 4 ellipticals) with $A_V{\,>\;}0.5$ have negative residuals.
The offset is significant at nearly the 3$\sigma$ level.
In order of increasing extinction, the 8~galaxies are E208-021,
NGC\,4976, E322-059, E092-013, E221-026, NGC\,3136, NGC\,2434, and NGC\,2380;
of these, only E322-059 has a positive residual.
There is no simple trend in the residuals at lower extinctions, however.
SFD produced a similar plot, showing residuals of the $(B{-}V)_0$--\mgii\
relation (data from Faber \etal\ 1989) against their $E(B{-}V)$ values.
Although they concluded that there was no significant overall trend, all
six galaxies with $E(B{-}V) > 0.2$ had negative residuals.
We note that Arce \& Goodman (1999) also found that the SFD map
over\-estimates the reddening in regions of smooth extinction 
with $A_V>0.5$ mag, although it may under\-estimate the reddening
in regions with steep extinction gradients.

Hudson (1999) has tested for large-scale systematic errors
in extinction maps by fitting dipoles to galaxy $(B{-}V)_0$--\mgii\
residuals as a function of position on the sky.  
He used a sample of 311 galaxies of high photometric quality 
from Faber \etal\ (1989) (colours within a radius of 33\arcsec)
and the homogenized
SMAC \mgii\ values and found weak evidence, at the 92\% confidence level,
for a dipole residual of 13\% amplitude in the SFD extinction map.
However, because the residuals with respect 
to a smaller, independent data set consisting
of Galactic globular clusters and RR~Lyra stars did not show this dipole,
Hudson concluded that the effect actually resulted from a 
$\sim\,$0.01 mag systematic error in the galaxy colours across the sky.

We have repeated Hudson's analysis, including the Monte Carlo estimations
of the significance, using our sample of 205 galaxies with
$\mgii>0.1$~mag.  Although the sample is smaller, \vi\ is $\sim\,$28\%
more sensitive to reddening errors.  We do not find a very significant
dipole residual.  The confidence level for the dipole term in our fit is
only 80\%, with fractional amplitude $0.09\pm0.07$. The best-fitting
direction, $(l,b) = (97^\circ,-5^\circ)$ in Galactic coordinates, is only
$30^\circ$ from that found by Hudson.  However, if just the four
highest extinction galaxies, all of which have negative residuals in
Figure~\ref{fig:mg2viext}, are omitted, then the fitted dipole swings by
more than 60$^\circ$ and has a similar low significance.  These four
highest extinction galaxies range in $l$ from 241$^\circ$ to 316$^\circ$
and in $b$ from $-$21\fdg5 to $+12\fdg9$.  We conclude that there is no
convincing dipole residual in the SFD extinction map from these data,
although the extinction values appear to be systematically overestimated
in certain directions.

\subsection{Environment}

We now explore the possibility of environmental influences on the \vimg\
relation.  Hudson (1999) found that cluster galaxies are $0.01\pm0.004$
mag redder in $(B{-}V)$ than group/field galaxies at a given \mgii.  Based
on stellar population models, this could imply that cluster ellipticals
are older and less enriched at a given \mgii, as might be expected in
hierarchical formation scenarios.  However, such an effect could also
imply environmental variations in the abundance ratios, presumably due to
differing time-scales for metal enrichment and the relative importance of
type I and type II supernovae, with shorter time-scales yielding a higher
relative abundance of alpha elements such as Mg (see Worthey 1998 for a
discussion).  In this case, one would expect cluster ellipticals to have
formed more rapidly and thus to have higher \mgii\ at a given \viz, or
equivalently, to be bluer at a given \mgii.

\begin{figure}
\medskip\vbox{\centering\leavevmode\hbox{
\epsfxsize=8.0cm\epsffile{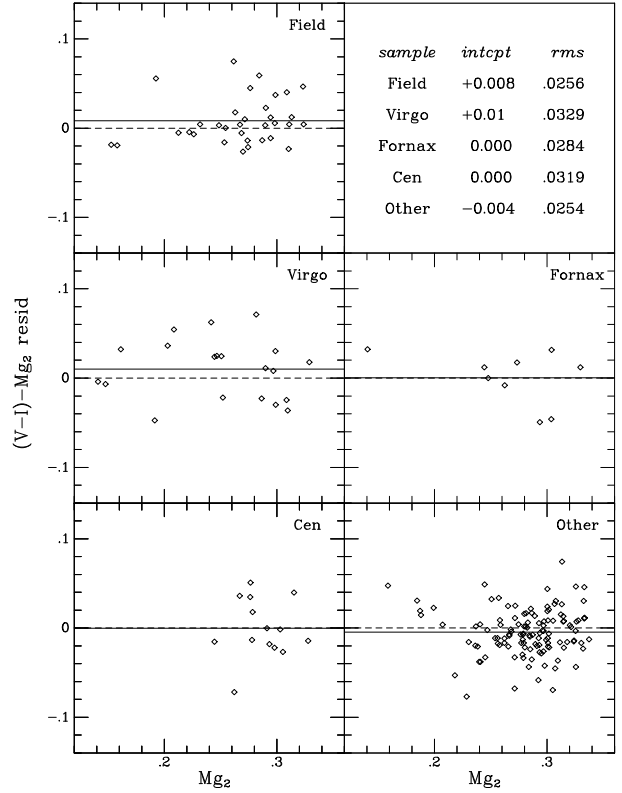}
}\caption{\small
Residuals from a linear fit to the \vimg\ relation are plotted against \mgii\
for galaxies in five different environments: the field, Virgo cluster,
Fornax cluster, Centaurus (A3526), and other small groups.
The vertical intercepts and rms residuals (mag) are listed at upper right.
\label{fig:mg2vienv}}}\medskip
\end{figure}

Figure~\ref{fig:mg2vienv} shows the residuals of the \vimg\
relation for galaxies in the field, in the three most prominent clusters,
and in other, smaller, groups.  The scatter among the Virgo galaxies is
the highest, but the sample is weighted more towards lower \mgii\ values
than the other samples and may be affected by a couple moderate outliers.
Formally, the offset in the \vimg\ relation between field and group/cluster
galaxies is $0.010\pm0.005$~mag, or $0.012\pm0.005$~mag if Virgo is omitted.
This is at the same $\sim2\sigma$ significance level as found by Hudson (1999),
but in the opposite sense: the field galaxies are marginally redder at
a given \mgii,
which would be consistent with longer enrichment time-scales.
However, at this level, systematic directional errors in the extinction
will become important, so we do not consider this result significant.

\subsection{Comparison to Stellar Population Models}
\label{ssec:heredity}

The Mg abundances of giant ellipticals are enhanced with respect to Fe
abundances, as compared to the solar ratios (e.g., Worthey, Faber, \&
Gonzalez 1992).  Variations in the Mg to Fe (logarithmic) ratio [Mg/Fe] at a
fixed total metallicity and age should cause scatter in the \vimg\ relation
just as it causes significant scatter in the plots of Fe indices against
Mg indices (e.g., Worthey 1998; Kuntschner 2000).  The effects of
non-solar abundance ratios on broad-band photometric colours due to
isochrone temperature and line blanketing effects is not well-modeled, so
we decided to look into this in a little more detail.

\begin{figure}
\medskip\vbox{\centering\leavevmode\hbox{
\epsfxsize=7.0cm\epsffile{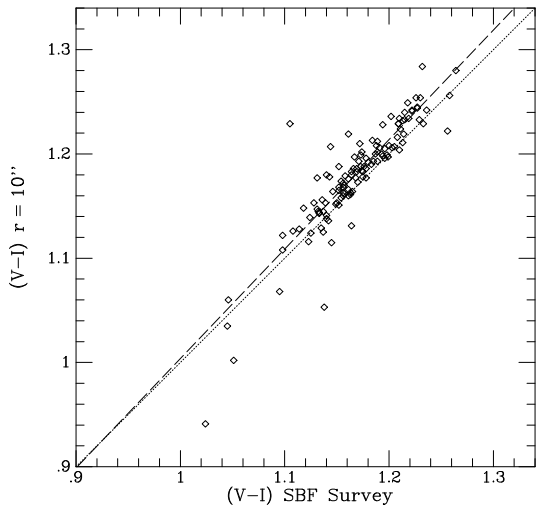}
}\caption{\small
Galaxy colour at a radius $r=10$ arcsec is plotted against colour from
the SBF survey data set for elliptical galaxies.  The dotted line is
the line of equality, and the dashed line is least-deviation fit to
the points.  SBF survey colours are typically $\sim0.015\,$mag
bluer than the colour at this fiducial radius.
\label{fig:sbfcenter}}}
\end{figure}

First it is necessary to make some correction for the systematic
difference in aperture between the \mgii\ and \viz\ measurements.
Most of our galaxies are at distances of $1300\pm400$ \kms\ 
(Virgo, Fornax, Leo, Dorado); at this distance the fiducial radius of 0.6 \hkpc\
corresponds to $\sim10$\arcsec.
We attempted to derive \viz\ colours near the galaxy
centres using the surface photometry data files described in SBF-I.
In many cases this was impossible because of saturation, for instance
the images from the CTIO runs (see SBF-I) with their large 0\farcs472
pixels and severe charge bleeding could not be used for this.
For the S0s, the central colours showed large scatter when plotted
against the SBF survey colours, owing to dust lanes and irregular
gradients associated with disk contamination, etc.  However, the relation
for the ellipticals was tighter; Figure~\ref{fig:sbfcenter} shows
\viz\ at $r=10\arcsec$ plotted against \viz\ from SBF-IV for the
ellipticals (regardless of whether or not they have \mgii\ from SMAC).
The least absolute deviation line (to reduce the effect of outliers)
shown in the figure is given by
\begin{equation}
(V{-}I)_0^{\rm ctr}\;=\; 1.05\,[\,(V{-}I)_0^{{\rm SBF}} -\, 0.92\,]\,,
	\label{eq:vigrad}
\end{equation}
where `ctr' actually refers to a radius of 10 arcsec.
The \viz\ aperture correction becomes larger for bigger, redder galaxies,
but is always less than 0.02~mag.

However, the \mgii\ value at $r=10\arcsec$ will be $\sim0.01$~mag
less than the value measured within this radius, according to the
mean gradient derived by Jorgensen \etal\ (1995).  Thus, we apply this
offset to the \mgii\ indices before making a comparison to the 
modified colours.  In addition, we follow Kuntschner \etal\ (2001) in
`correcting' the \mgii\ indices to what they would be for 
solar abundance ratios at the same total metallicity~$Z$.  We use the
results from their sample of 72 early-type galaxies to derive
\begin{equation}
\mgii^{\rm sol} \;=\; 0.022 + 0.87\,\mgii\,,
	\label{eq:mg2solar}
\end{equation}
where $\mgii^{\rm sol}$ is the \mgii\ value corrected to solar-abundance
ratios.  The scatter in this relation is 0.005~mag.

\begin{figure}
\medskip\vbox{\centering\leavevmode\hbox{
\epsfxsize=8.0cm\epsffile{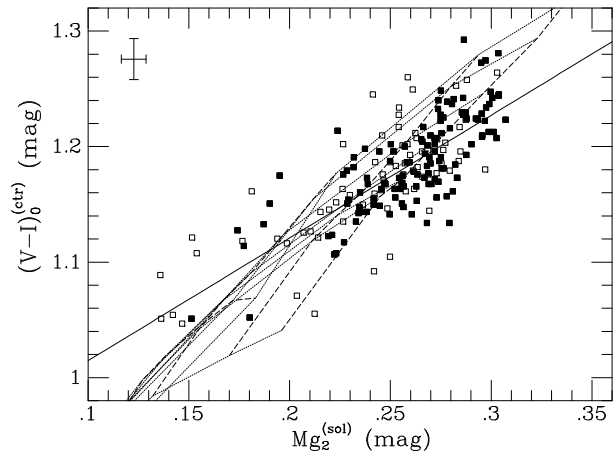}
}\caption{\small
\viz\ colour, corrected to the `central' value, is plotted
against \mgii, corrected to solar abundance ratios following
Kuntschner \etal\ (2001).  Filled and open squares are for
ellipticals and S0s, respectively. The grid shows updated Vazdekis (1996)
stellar population models from Blakeslee, Vazdekis, \& Ajhar (2001),
with metallicities in the range $\hbox{[Z/H]} = -1.7$ to $+0.2$ dex
and ages from 2 to 18~Gyr.
The solid line is a fit to the filled points; it has a slope of 1.06, as
compared to 0.88 for the corresponding fit in Figure~\ref{fig:mg2vimain}.
However, the models define a yet steeper slope of $\sim\,$1.5.
\label{fig:vimg2mod}}}
\end{figure}

Figure~\ref{fig:vimg2mod} shows the modified \viz\ and \mgii\ values as
compared to some of the latest (solar abundance ratio) stellar population
models from Vazdekis (2001, in preparation; see Blakeslee, Vazdekis,
\& Ajhar 2001), based on extensive empirical stellar libraries.
It is clear that the data do not follow the model locus.
In particular, the galaxies with corrected $\mgii^{\rm sol} > 0.25$
have colours which are still too blue with respect to the models.
The apparent offset in the other direction for $\mgii^{\rm sol}< 0.2$
is less worrisome, as this region is less well-populated by
nice ellipticals with well-defined gradients, and only one of the 72
Kuntschner \etal\ (2001) galaxies which went into deriving
Eq.\,(\ref{eq:mg2solar}) had $\mgii<0.20$.  If the $\sim {-}0.03$ mag
offset in the colours at high $\mgii^{\rm sol}$ is due to non-solar
abundance ratios, then \viz\ colour for early-type galaxies must increase,
just as the Fe indices increase (see Trager \etal\ 2000
and Kuntschner \etal\ 2001), when corrected to solar ratios
(i.e., lower Mg at fixed~$Z$).

There has been much discussion as to whether it is more proper to call Mg
and the other alpha elements `enhanced' or the Fe-peak elements
`depressed' in giant ellipticals (e.g., Vazdekis \etal\ 1997; Trager
\etal\ 2000, Kuntschner \etal\ 2001), but both are relative terms.  In any
case, our results indicate that \viz\ colour is not `enhanced' like Mg in
these galaxies, but rather is `depressed' towards the blue, consistent
with the results of Salaris \& Weiss (1998)
who found that isochrones with $\hbox{[Mg/Fe]}>0$ were hotter, and thus
bluer, at all stages in the stellar evolution than scaled-solar abundance
isochrones at a given total metallicity.  The difference in temperature
became much more pronounced for the higher-metallicity isochrones.

\subsection{Distance-Independent Mass Measures}

SBF-IV introduced the distance-independent `fluctuation star count'
\Nbar, defined as the difference between the SBF magnitude \mbar\ and the
total galaxy magnitude:
\begin{equation}
\Nbar \,=\, \mbar - m_{\rm tot} \,=\,
    +2.5\,\log\left[L_{\rm tot} \over \lbar\right] \,,
\label{eq:nbardef}
\end{equation}
where $L_{\rm tot}$ is the total luminosity and \lbar\ is the fluctuation
luminosity (the luminosity-weighted mean stellar luminosity).  Therefore,
\Nbar\ corresponds to 2.5 times the logarithm of the luminosity-weighted
number of stars in a galaxy; as such, it essentially scales with the
galaxy mass.  Because \mbar\ and $m_{\rm tot}$ are measured in the same
bandpass, this parameter is independent of Galactic extinction and, if
both are from the same image, independent of photometric zero~point. 

\begin{figure}
\medskip\vbox{\centering\leavevmode\hbox{
\epsfxsize=7.8cm\epsffile{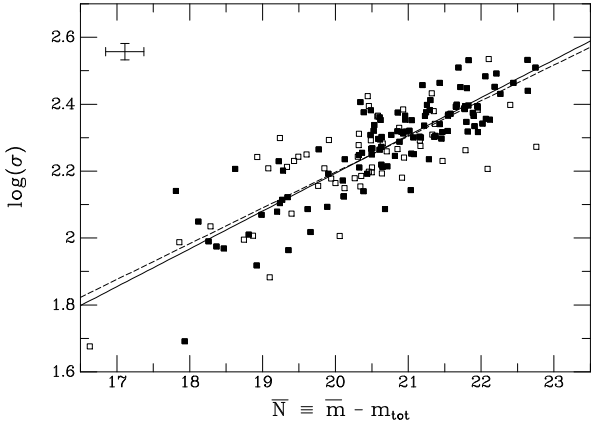}
}\caption{\small
The stellar velocity dispersion $\sigma$ from the SMAC catalogue
is plotted against the `fluctuation star count' parameter \Nbar\
discussed by SBF-IV.  Filled and open circles are ellipticals
and S0s, respectively; the dashed line is a fit to the full sample,
and the solid line is a fit the ellipticals.
\label{fig:Nbar_sig}}}\medskip
\end{figure}

SBF-IV showed that \Nbar\ correlated well with the galaxy
velocity dispersion for a sample of 64 ellipticals with
velocity dispersions from Prugniel \& Simien (1996).  
Figure~\ref{fig:Nbar_sig} shows the corresponding plot
for 174 SBF survey galaxies with $m_{\rm tot}$ values
from the fits described in the following section
and SMAC velocity dispersions. The least-squares fit
to the ellipticals and S0s with $\Nbar>18$ is
\begin{equation}
\log\sigma \;=\; 2.20 \,+\, 0.10\,(\Nbar - 20)\,,
	\label{eq:nbar_sig}
\end{equation}
with an rms scatter of 0.079~dex; the 117 ellipticals give
a closely consistent result and have a scatter of 0.069~dex.
Thus, the relation is even tighter than the $\log\sigma$--\mgii\
relation, which has a scatter of 0.084~dex for the same sample
of galaxies (both the full sample and elliptical subsample).

Eq.\,(\ref{eq:nbar_sig}) is nearly identical to the relation 
found by SBF-IV for their smaller sample.  Rearranging terms,
we have
\begin{equation}
 L_{\rm tot} \;=\;  10^8 \lbar \,
\left(\sigma \over 160{\,\rm km\, s^{-1}}\right)^{4.0} \,,
\label{eq:fabernbar}
\end{equation}
which resembles the Faber-Jackson relation, except
that the \lbar\ term has a dependence on mass-to-light ratio that results
in the small scatter for Eq.\,(\ref{eq:nbar_sig}).  Blakeslee \etal\
(2001) discuss the stellar-population aspects of \Nbar\ in more detail.
We note that the value of the exponent in Eq.\,(\ref{eq:fabernbar})
depends on the inverse nature of
our least-squares fit. A `forward' fit of \Nbar\ as a function of
$\log\sigma$ would have given an exponent of about~2.8; ideally, one
would do a full maximum-likelihood analysis.  
The use of \Nbar\ for an alternative calibration of the SBF
method is discussed in Paper~II.

\section{FP Photometric Parameters from SBF Data}
\label{sec:fpderiv}

Relatively few of the SBF survey galaxies with SMAC spectroscopic
parameters also have SMAC surface photometry.  This is largely because the
SMAC project did not use the surface photometry from the 7S survey
(Burstein \etal\ 1987), which is in the $B$-band and uncorrected for
seeing effects, while it did include the recalibrated and merged 7S
spectroscopic data (Davies \etal\ 1987).  Therefore, we have derived FP
photometric parameters from SBF survey data. This presented a number of
challenges, as the field sizes were not ideal and the galaxy centres often
saturated.  However, SBF-I described a procedure for measuring surface
photometry in a series of concentric circular annuli of 3\arcsec\ width
for all the SBF survey images after masking and interpolating over the
bright stars.  The surface photometry data were used in SBF-I to determine
accurate photometric offsets for the many individual observing runs in the
SBF survey and to provide accurate calibrations for non-photometric survey
data.

We fitted Sersic (1968) $r^{1/n}$ profiles to the SBF survey surface
photometry data in circular annuli.  A generalization of the
de\,Vaucouleurs (1948) law, the $r^{1/n}$ profile is convenient because it
describes a large range of galaxy types from exponential disks to highly
diffuse cD halos.
[See Ciotti \& Lanzoni (1997) and Ciotti \& Bertin (1999) for discussions
of the general properties of stellar systems following $r^{1/n}$ profiles.]
This profile has been widely
used for empirically characterizing galaxy light distributions (e.g.,
Davies \etal\ 1988; Young \& Currie 1994; Caon, Capaccioli \& D'Onofrio
1993; Courteau \etal\ 1996), including specifically for measuring FP
distances (Graham \& Colless 1997; Graham 1998).  Kelson \etal\ (2000)
also explored $r^{1/n}$-law fits and concluded that the difference with
respect to the $r^{1/4}$-law fits was negligible for their subset of SBF
survey data; however, the full data set is much more morphologically
diverse, so we have kept this extra degree of freedom.

Not all the of the SBF survey galaxies converged to reasonable $r^{1/n}$
fits (reduced $\chi^2 \lta 1$).  The failures to produce a good fit were
partly because of morphological irregularity (9\% of the 299 galaxies in
the sample presented by SBF-IV were classified as spirals) and partly
because of problems in the data (central saturation, bright nearby stars,
occasionally poor flattening, etc.) that could be overcome for the SBF
analysis, but not for the surface photometry fitting.  We fitted the $V$
and $I$ SBF surface photometry down to fixed fractions of the sky
brightness, 2.0\% in $V$ and 0.4\% in $I$.  Acceptable $r^{1/n}$ fits were
obtained for a total of 280 galaxies, of which 268 were elliptical or
lenticular ($T\le0$); 257 of these were successfully fitted in both the
$V$ and $I$ bands.
The fits were integrated to find the total magnitudes, half light (or
effective) radii $R_e$, and mean surface brightnesses \sbe\ interior to
the half light radius.

The surface brightnesses are corrected for Galactic extinction from SFD
and $(1{+}z)^4$ dimming in all cases.  We use the same $K$-corrections as
in SBF-I: $1.9z$ for $V$ magnitudes and $1.0z$
for $I$ magnitudes.  Because the SMAC survey data are standardised to
the $R$-band, we transform the $V$ and $I$ surface brightnesses to $R$ according
to
\begin{equation}
 (V{-}R)_0 \;=\; 0.034 \,+\, 0.46\,(V{-}I)\,, \label{eq:vrvirel}
\end{equation}
which comes from the empirically-based models discussed in
\S\ref{ssec:heredity}.  At the median colour of $\viz=1.16$, this implies
$(V{-}R)_0 = 0.57$, identical to the fixed colour transformation used by
Smith \etal\ (1997) and only 0.01~mag different from the fixed color
assumed by SMAC-III.
As a further check, we derived a rough transformation based on data for
11 early-type galaxies with $VRI$ colours measured by Tonry, Ajhar, \&
Luppino (1990), whose \viz\ colours span a range of 0.20 mag and share a
homogeneous system with our data.  The resulting relation has a slope
of $0.42\pm0.06$, consistent with Eq.\,(\ref{eq:vrvirel}), and is bluer in
the mean by $0.015\pm0.004$ mag.  This $(V{-}R)$ zero-point offset is typical
of the \vi\ offsets among SBF survey runs prior to homogenization (SBF-II)
and is in the opposite sense to the 0.01 mag offset with respect to the
SMAC-III colour.  We conclude that Eq.\,(\ref{eq:vrvirel}) is accurate at
the 0.01--0.02 mag level.

Finally, we form $\xfp\equiv\log R_e - 0.33\sbe$,
which is the photometric combination entering into the FP.
We have neglected seeing corrections on \xfp\ for this sample of nearby
galaxies, all of which have $10\arcsec\lta R_e\lta 100\arcsec$
and were observed in a typical seeing of 1\arcsec.

Table~1 presents the results of our fits to the SBF survey photometry.
The colums list: (1) galaxy name and (2) heliocentric velocity as in
SBF-IV; (3) morphological $T$-type from the RC3; (4) $B$-band extinction
from SFD; (5) \viz\ colour from the SBF survey; (6) \Nbar\ fluctuation
count parameter formed from the SBF survey \mi\ and our fitted $I$-band
$m_{\rm tot}$; (7) \PD\ quality parameter defined in SBF-II and tabulated in
SBF-IV; (8) $\log R_e$ from the $I$-band fit; (9) mean effective $R$-band
surface brightness \sbeR\ from the $I$-band fit; 
(10) $\log R_e$ from the $V$-band fit; (11) \sbeR\ from the $V$-band fit;
(12) $\log\sigma$ from the SMAC survey; and (13) \mgii\ from the SMAC survey.
The surface brightnesses are corrected for Galactic extinction
and $(1{+}z)^4$ dimming, $K$-corrected, and transformed to the $R$-band
as described above.  The errors on \viz,
$\log\sigma$, and \mgii\ are taken from the SBF and SMAC catalogues.  For
the \Nbar\ error, we added the \mi\ error in quadrature with a nominal
0.2~mag uncertainty in $m_{\rm tot}$.  We have included the \PD\
parameter for further information on the quality of the
\mi\ measurement, with high values ($\PD\gta2.7$) indicating poor quality
(see SBF-II and Paper~II for discussions).  We have not included error
estimates on the $\log R_e$ and \sbe\ values, as they are too strongly
correlated, and we recommend using them only in the \xfp\
combination.  Most of the remainder of this paper is dedicated to
evaluating the uncertainty in the \xfp\ values.

\begin{figure}
\medskip\vbox{\centering\leavevmode\hbox{
\epsfxsize=8.0cm\epsffile{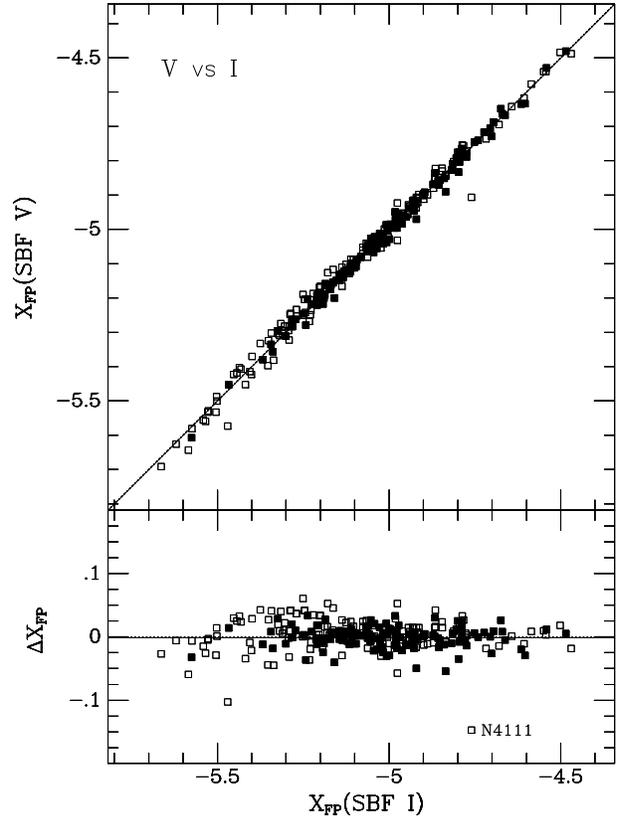}
}\caption{\small
Comparison of the FP photometric parameter $\xfp \equiv \log R_e - 0.33\sbe$ 
determined from $V$-band and $I$-band SBF survey surface photometry
transformed to the $R$-band for comparison.
The top panel is a direct comparison of \xfp\ for the early-type
galaxies, and the lower panel shows the residuals
$\Delta = \xfp({{\rm SBF}\,V}) -  \xfp({{\rm SBF}\,I})$.
Open and filled symbols represent S0s and ellipticals, respectively.
The dotted lines show equality and the solid lines are fits with
unit (top) and zero (bottom) slopes.  The overall scatter decreases
from 0.023 to 0.021~dex if the edge-on S0 NGC\,4111 is omitted; the
scatter for just the ellipticals is 0.016~dex.
\label{fig:XFPsbfv_sbfi}}}\medskip
\end{figure}

Figure~\ref{fig:XFPsbfv_sbfi} shows the comparison between the $R$-band
\xfp\ values determined from the $V$ and $I$ SBF photometry.  The overall
scatter is 0.023~dex for 257 early-type galaxies.  The largest outlier, at
0.014~dex, is NGC\,4111, which is a nearly edge-on (axis ratio $>5$) S0
galaxy with a red bulge and bluer disk.  It does not appear in the 7S or
SMAC catalogues and therefore is not used for the distance comparisons in
Paper~II.  Excluding NGC\,4111, the scatter in Figure~\ref{fig:XFPsbfv_sbfi}
is 0.021~dex.  For the 122 ellipticals ($T{\,=\,}{-}5$), the
scatter is 0.016~dex.  The internal error per measurements is
therefore about 0.015~dex overall, or 0.011~dex for ellipticals.
Table~\ref{tab:xfpres} summarizes the results of this comparison
and the external comparisons below.

\section{External FP Comparisons}

\begin{figure}
\medskip\vbox{\centering\leavevmode\hbox{
\epsfxsize=8.0cm\epsffile{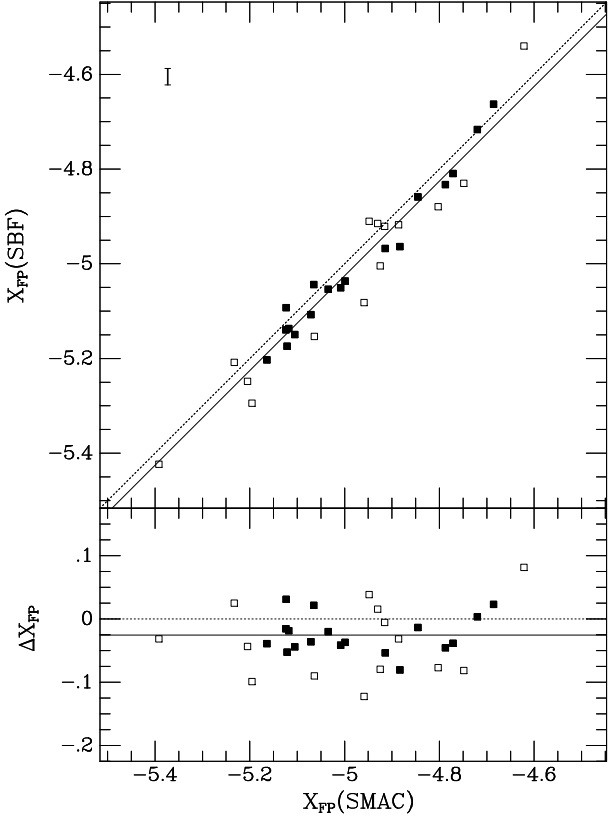}
}\caption{\small
Comparison of the FP photometric parameter $\xfp \equiv \log R_e - 0.33\sbe$ 
determined from the $I$-band SBF survey surface photometry (transformed approximately
to the $R$~band before calculating \xfp)
with \xfp\ values determined from the SMAC survey dataset.  Top panel is
a direct comparison and the lower panel shows the residuals
$\Delta = \xfp({\rm SBF}) -  \xfp({\rm SMAC})$.
Open and filled symbols represent S0s and ellipticals, respectively.
The dotted lines show equality and the solid lines are fits with
unit (top) and zero (bottom) slopes.  A correction of $\sim0.03$~dex 
is required in order to transform the \xfp(SBF) values to
the homogeneous SMAC FP system.
\label{fig:XFPsmacsbfi}}}\medskip
\end{figure}

\begin{figure}
\medskip\vbox{\centering\leavevmode\hbox{
\epsfxsize=8.0cm\epsffile{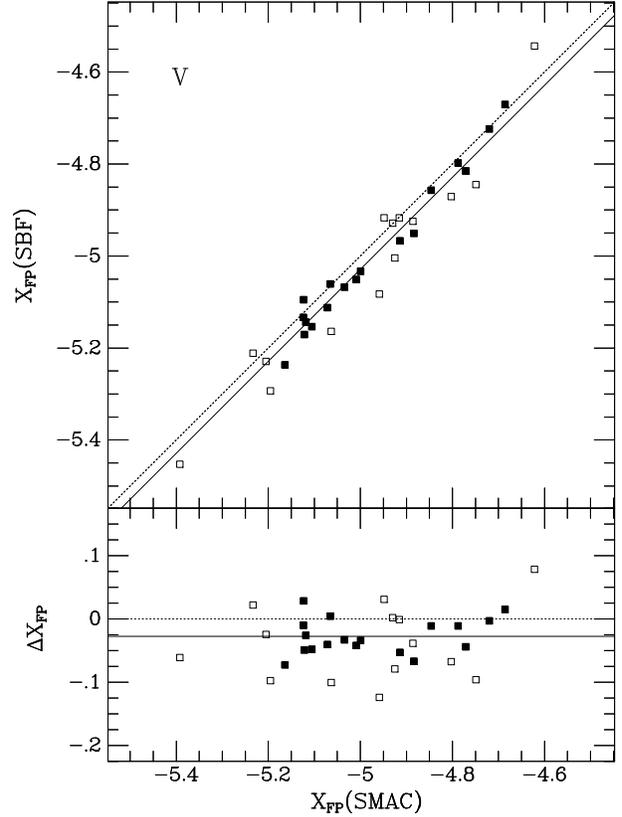}
}\caption{\small
Comparison of the FP photometric parameter $\xfp \equiv \log R_e-0.33\sbe$ 
determined from the $V$-band SBF survey surface photometry (transformed
approximately to the $R$~band before calculating \xfp) with \xfp\ values
determined from the SMAC survey dataset.  Top panel is a direct comparison
and the lower panel shows the residuals $\Delta = \xfp({\rm SBF}) -
\xfp({\rm SMAC})$.  Symbols and lines are as in
Figure~\ref{fig:XFPsmacsbfi}.  Again, a correction of $\sim0.03$~dex is
required in order to transform the \xfp(SBF) values to the homogeneous
SMAC FP system.
\label{fig:XFPsmacsbfv}}}\medskip
\end{figure}

Figures~\ref{fig:XFPsmacsbfi} and~\ref{fig:XFPsmacsbfv} compare \xfp\ from
the $I$ and $V$ SBF surface photometry fits of the previous section to
\xfp\ from the SMAC survey.  There are 32 galaxies, including 18
ellipticals, with \xfp\ from both data sets.  Although our $V$ and $I$
fits were done independently of each other, these figures look remarkably
similar in detail, not just in their offsets and scatter.  This 
is because the internal scatter is significantly smaller than
the external scatter, which has contributions from systematic effects
resulting from the different fitting techniques, etc.  The biggest
outlier in both figures is the Dorado S0 galaxy NGC\,1553, which also has
the largest \xfp\ in these figures.  The data in the SMAC catalogue for
this galaxy derive from the Jorgensen \etal\ (1996) study.

As shown in Table~\ref{tab:xfpres},
the scatter in \xfp\ for the SMAC--SBF survey
comparisons is 0.044~dex for the full overlap sample, 0.039--0.040 dex if
NGC\,1553 is omitted, and 0.027--0.029 dex for just the ellipticals.
The offset between the two sets of \xfp\ values is about $0.030\pm0.007$ dex,
in the sense of the SMAC values being larger. In comparison, SMAC-III
did not find significant \xfp\ offsets among the different studies
which went into the SMAC catalogue.  The offset found here is not due
to the photometry, as we directly compared photometric zero points
for 10 of the 32 galaxies in common (all having new photometry from
the SMAC survey), and the systematic offset was $0.03\pm0.01$ mag,
a factor of three too small.  Therefore, the \xfp\ offset must
come from the difference in fitting procedures used and may
partially reflect the lack of seeing corrections.  However,
the agreement is reasonable after the application of this systematic
correction; there is no evidence for systematic trends in the residuals
with \xfp, redshift, or galaxy type (although the scatter is larger
for the S0s).

We can use the 7S photometric data set as an additional external 
check of our \xfp\ measurements.  We downloaded the data from
Faber \etal\ (1989) from the Astronomical Data Center
and calculated \xfp\ values from the $B$-band surface brightnesses,
$(B{-}V)$ colours, and effective diameters.  
The Burstein \& Heiles (1984) extinction corrections were removed
and the SFD extinction corrections were applied.
The surface brightnesses were corrected for $(1{+}z)^4$ dimming
(the 7S catalogue incorrectly states that this was already done),
and then transformed to the $R$-band using
\begin{equation}
 (V{-}R)_0 \;=\; 0.126 \,+\, 0.48(B{-}V)_0\,. \label{eq:vrbvrel}
\end{equation}
Since this is an extrapolation of the colours, it is more uncertain than
Eq.\,(\ref{eq:vrvirel}), but the range of galaxy colours is small enough
that the typical error in $(B{-}R)_0$, and hence the $R$-band surface
brightness, should only be a few hundredths of a magnitude, or $\sim0.01$
dex in \xfp. 

\begin{figure}
\medskip\vbox{\centering\leavevmode\hbox{
\epsfxsize=8.0cm\epsffile{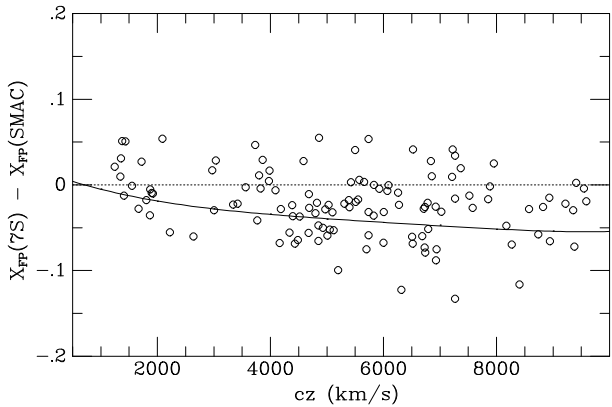}
}\caption{\small
Differences in the \xfp\ parameters between the SMAC and 7S surveys
are plotted against heliocentric velocity, used as an indicator of distance.
The 7S values have been transformed to the $R$~band with SFD extinction
corrections and corrected for $(1{+}z)^4$ surface brightness dimming.
The systematically negative residuals at $c{z}\gta5000$ \kms\ result
from the lack of seeing corrections in the 7S data.
The solid curve illustrates example SMAC seeing corrections 
for a galaxy with $R_e = 1.6$~\hkpc\ observed in 2\arcsec\ seeing
at different redshifts.
\label{fig:dXFPsmac7s}}}\medskip
\end{figure}

We noted above that the 7S photometric data were omitted from the SMAC
survey because they were not corrected for seeing effects.
Figure~\ref{fig:dXFPsmac7s} plots the difference in the \xfp\ values
between the SMAC and 7S catalogues against redshift
and compares an example SMAC seeing correction
curve for a galaxy of fixed physical size $R_e=1.6$ \hkpc\
(5\arcsec\ at Coma distance) observed in 2\arcsec\ seeing.
This model appears to overestimate the correction nearby, but gives
a good match to the offset at larger distances.
Table~\ref{tab:xfpres} includes the results of the 7S-SMAC comparison for
galaxies within $c{z}<5000$ \kms, for which the observed offset is not
significant.
All of the SBF survey galaxies have $c{z}<5000$ \kms\ (and only a
few have $c{z}>4000$ \kms); this is what has allowed us to
neglect seeing effects in our own analysis.
Before proceeding with the comparisons however, we note that \xfp\
disagreed for NGC\,1389 by more than 0.45~dex between the SBF and 7S
data~sets (it is not in the SMAC photometric catalogue).  The $\log R_e$
values agreed to within 0.04\,dex, but the transformed 7S \sbe\ is fainter
by 1.44 mag.  This galaxy has `quality~3' (the lowest) photometry in the
7S data set, but other quality~3 galaxies do not show similarly large
disagreements, so we omit NGC\,1389 only from the comparisons below.

\begin{figure}
\medskip\vbox{\centering\leavevmode\hbox{
\epsfxsize=8.0cm\epsffile{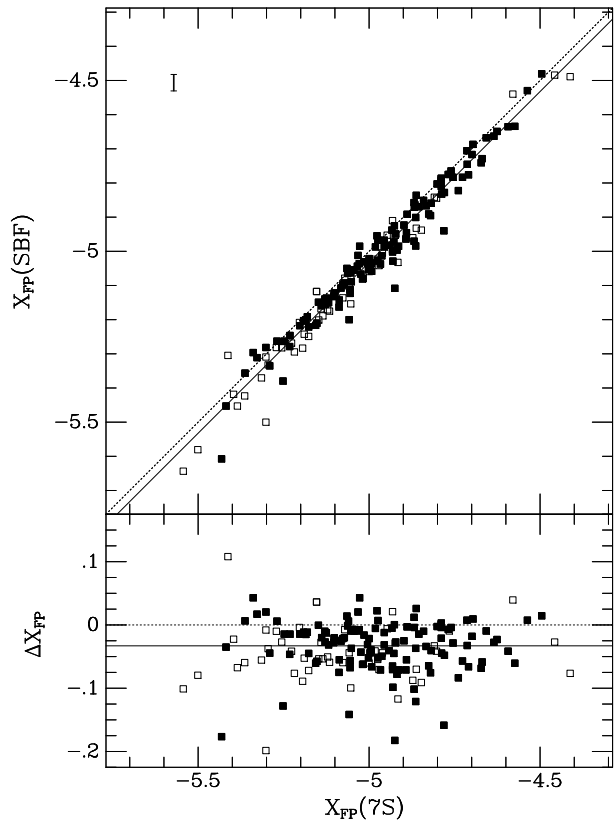}
}\caption{\small
Same as Figure~\ref{fig:XFPsmacsbfi} but for the comparison between \xfp\
determined from $I$-band SBF survey data and \xfp\ determined from
the 7S survey dataset of Faber \etal\ (1989) (both transformed to $R$ band).
\label{fig:XFPsbfi7s}}}\medskip
\end{figure}

\begin{figure}
\medskip\vbox{\centering\leavevmode\hbox{
\epsfxsize=8.0cm\epsffile{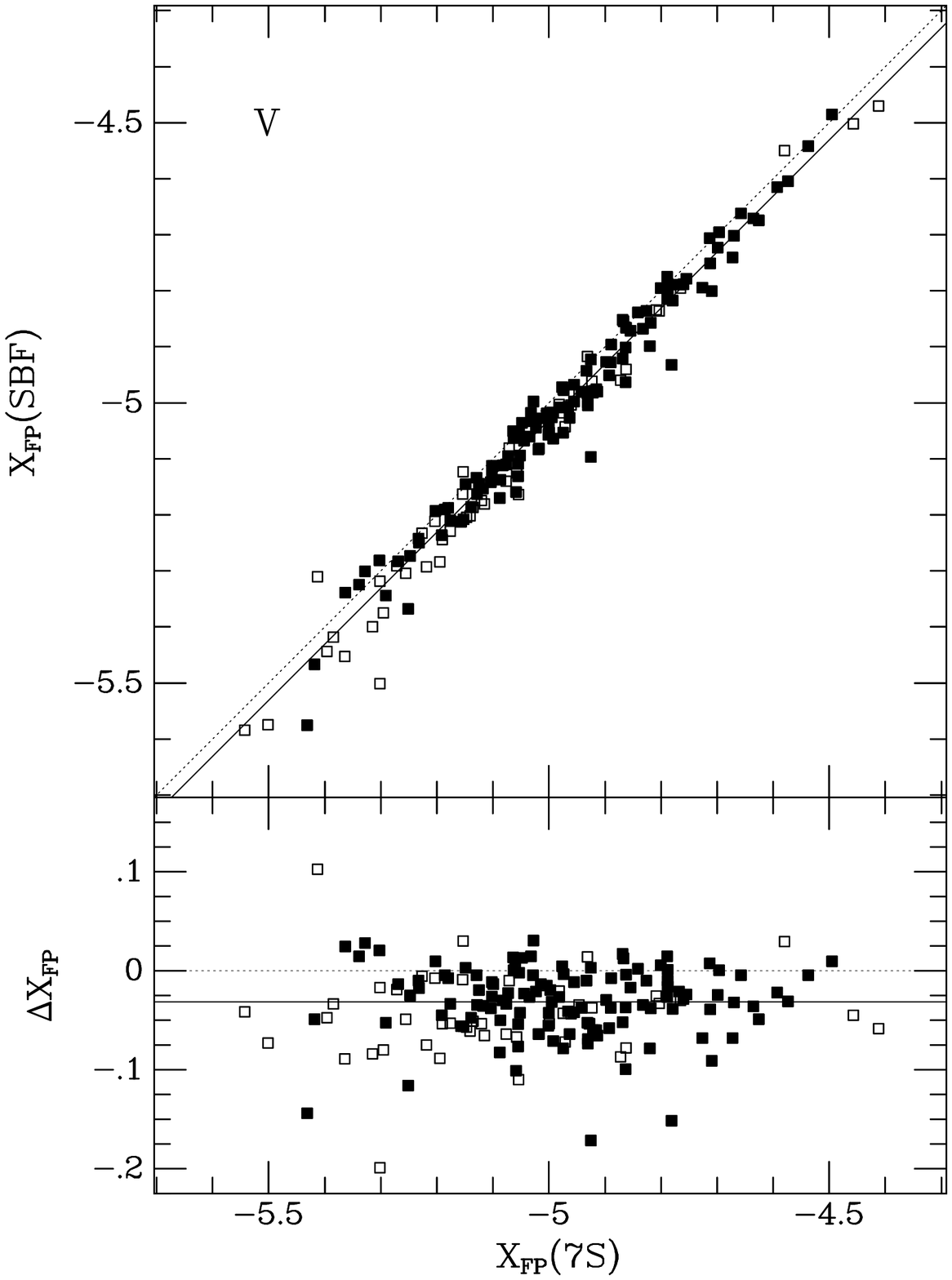}
}\caption{\small
Same as Figure~\ref{fig:XFPsmacsbfv} but for the comparison between \xfp\
determined from $V$-band SBF survey data and \xfp\ determined from
the 7S survey dataset of Faber \etal\ (1989) (both transformed to $R$ band).
\label{fig:XFPsbfv7s}}}\medskip
\end{figure}

\begin{figure}
\medskip\vbox{\centering\leavevmode\hbox{
\epsfxsize=7.7cm\epsffile{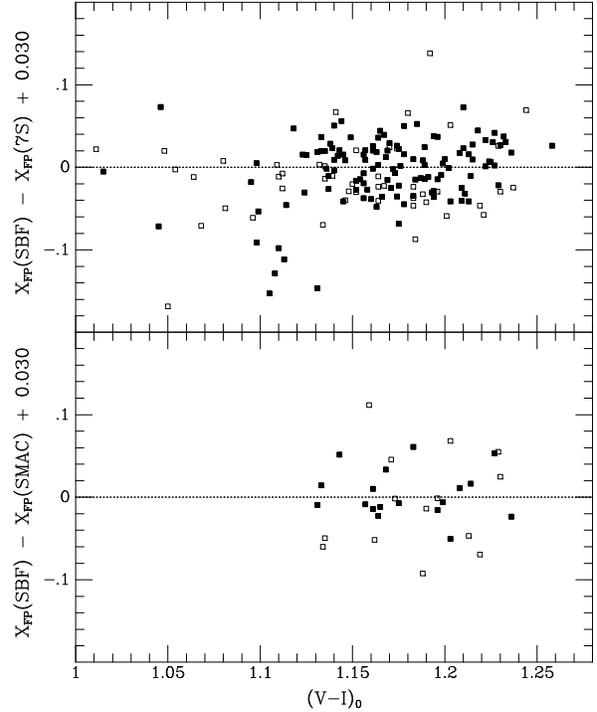}
}\caption{\small
Differences in the \xfp\ values between the 7S and $I\,$SBF 
dat sets (top) and between the SMAC and $I\,$SBF data sets (bottom)
are plotted against \viz\ from the SBF survey.
\label{fig:dXFP_VI}}}\medskip
\end{figure}

Figures~\ref{fig:XFPsbfi7s} and~\ref{fig:XFPsbfv7s} show the comparisons
of the transformed $R$-band SBF and 7S \xfp\ values. The results are again
summarized in Table~\ref{tab:xfpres}. There is no trend of the residuals
with redshift.  However, the galaxies which lie significantly below the
mean difference line for these SBF-7S comparisons are bluer than average.
Figure~\ref{fig:dXFP_VI} plots the differences against \viz\
colour. Galaxies with $\viz\lta1.13$ exhibit greater scatter and tend to
have 7S \xfp\ values greater than expected from the comparison to the SBF
survey photometry.  Table~\ref{tab:xfpres} shows that the scatter
decreases by $\sim\,$25\% when only $\vi\ge1.135$ galaxies are considered.
This could be due to stellar population effects, as the 7S photometry
comes from the $B$-band, while the SBF photometry is in $V$ and $I$, and
both have been transformed to $R$.  However, it is not a simple linear
relation between the residuals and colour (there is no trend for the redder
galaxies which constitute the bulk of the sample), and photometric
transformation errors should not cause offsets in \xfp\ as large as
0.1~dex.  We cannot test for a similar problem in the SMAC comparison
because the overlapping galaxies all have $\viz>1.13$ (lower panel of
Figure~\ref{fig:dXFP_VI}).  We use the \xfp\ values from the SBF survey
data and test for colour-dependent errors in the distance analyses of
Paper~II.

For the purpose of comparison to other works, the last column of
Table~\ref{tab:xfpres} lists the rms scatters in comparisons of \xfp\
defined with a \sbe~coef\-ficient of 0.30 instead of 0.33. 
For instance, Smith \etal\ (1997) used a coefficient of 0.30 in doing
these sorts of \xfp\ comparisons, although the actual calculation of FP
distances used a coefficient of 0.326 (Hudson \etal\ 1997).  
As Table~\ref{tab:xfpres} shows, the magnitude of the scatter is
reduced by 20--35\% for \xfp\ comparisons using the smaller coefficient
(the scatter would increase by a similar amount if a coefficient of
0.36 were used).  The systematic offset of our \xfp\ values would
change from $\sim\,$0.030 to $\sim\,$0.020 dex for a \sbe\ coefficient
of 0.30 (and to $\sim\,$0.040 dex for a coefficient of 0.36).
We have adopted 0.33 for the \sbe\ coefficient in the FP relation because
it is within 0.01 of the values derived from very large data samples by
Hudson \etal\ (1997), Jorgensen \etal\ (1996), Colless \etal\ (2001) and
the SMAC project.

From the various comparisons, we can formally derive the error on \xfp\
for any of the data sets.  Specifically, we find that the external errors
on our measurements of \xfp\ from SBF survey data are 0.030-0.035~dex, or
0.020--0.025~dex if just the ellipticals are considered, with the 7S \xfp\
errors being about 0.030~dex for both ellipticals and S0s.  The \xfp\ errors
translate to an error of 6--7\% in distance, roughly a factor of 3 smaller
than the error expected from the intrinsic scatter in the FP.  Thus, our
\xfp\ measurements are accurate enough for measuring FP distances, and the
systematic correction of 0.03~dex brings them onto the homogeneous SMAC
system.

\section{Summary}

We have merged the SBF survey data with the SMAC FP survey data and
explored the correlations between the distance-independent galaxy
properties measured in the two surveys.  Using 200 galaxies with data
in both catalogues, we derived mean relations between SBF survey \viz\
colour and SMAC \mgii\ and velocity dispersions.  The intrinsic scatter in
the \vimg\ relation (including the effects of colour gradients and any
reddening errors) is 0.020~mag for the ellipticals and
0.022~mag for the S0s.  We used this relation to test for errors in
the SFD reddenings.  There is no convincing evidence 
from these data for a dipole residual in the SFD map
(our best dipole fit has a 9\% amplitude but only 80\% confidence).
However, we do find that the galaxies in our sample with SFD
$A_V\gta0.55$ mag have overestimated extinctions.  Because these galaxies
lie near the Galactic plane in a fairly restricted longitude range, we
cannot say whether or not the problem is systematic at high extinction
or simply indicative of large-scale coherent errors.

We also tested for the influences of environment and $\alpha$-element
enhancement on the \vimg\ relation.
Consistent with expectations from the $\alpha$-enhanced isochrone
calculations of Salaris \& Weiss (1998), we found that the enhancement of
the \mgii\ values is not mirrored by redder \viz\ colours.  In fact,
comparison to solar-abundance ratio stellar population models indicates
that \viz\ is slightly `depressed' at a given total metalicity, and
therefore appears to follow the Fe-peak abundances more than 
the $\alpha$-elements abundances.  We find no evidence for environmental
effects larger than $0.01\pm0.005$ mag on the \vimg\ relation. 

We have derived the relation between the stellar velocity dispersion
$\sigma$ from the SMAC catalogue and the `fluctuation star count' \Nbar,
which was introduced in SBF-IV and corresponds to the
luminosity-weighted number of stars in a stellar population.  
The scatter is only 0.079~dex in $\log\sigma$ (0.069~dex for the
ellipticals), making this even tighter than the \mgii--$\log\sigma$
relation.  The \Nbar-$\log\sigma$ relation is equivalent to the
Faber-Jackson relation with a stellar-population correction term.

Finally, we have used the SBF survey $V$- and $I$-band photometry to
derive the FP photometric parameters \xfp\ for the survey galaxies.
Comparisons to the SMAC and 7S data samples allowed us to estimate the
external errors on our \xfp\ measurements.  A systematic correction of
about 0.03~dex is required to bring our \xfp\ values onto the SMAC
photometric system.  We provide a full data table of our results, to be
used in a forthcoming analysis paper comparing SBF and FP distances to
these galaxies.~~~

\section*{Acknowledgments}
We thank our SBF and SMAC survey collaborators Ed Ajhar, Roger Davies,
Alan Dressler, David Schlegel, and Russell Smith for their help in
amassing and processing these data~sets.  We are grateful to Harald
Kuntschner for providing his \mgii\ data in electronic format.  This work
was supported at the University of Durham by a PPARC rolling grant in
Extragalactic Astronomy and Cosmology and made use of Starlink computer
facilities.  JPB thanks the ACS project at Johns Hopkins University for
support while finishing this paper.


\begin{table*}\setcounter{table}{0}
\centering \begin{minipage}{172mm}
\caption{Fundamental Plane Data for SBF Survey Galaxies}\label{tab:data}
\tabcolsep=0.14cm\footnotesize
\begin{tabular}{lrrcccccccccc}
\noalign{\vspace{3pt}\hrule\vspace{3pt}}
Galaxy & $v_h\;$ & $T$ & $A_B$ & \viz~ $\pm$~ & ~\Nbar~~~ $\pm$ & \PD\ &
 $\log R_e^{(I)}$ & $\langle \mu_R \rangle_e^{(I)}$ & $\log R_e^{(V)}$ & $\langle \mu_R \rangle_e^{(V)}$ &
 $\log\sigma$~~ $\pm$~~ & Mg$_2$~~~ $\pm$~ \\
\noalign{\vspace{3pt}\hrule\vspace{3pt}}
E092-013 & 1770 &$-$4 & 0.811 & 1.164~ 0.033 & 19.43~ 0.23 & 1.52 & 1.306 & 19.621 & 1.448 & 20.150 &  2.230~ 0.029 & 0.277~ 0.008 \\
E183-030 & 2668 &$-$3 & 0.376 & 1.139~ 0.014 & 20.94~ 0.57 & 3.14 & 1.375 & 19.483 & 1.408 & 19.491 &  2.241~ 0.042 & 0.256~ 0.010 \\
E208-021 & 1003 &$-$3 & 0.750 & 1.146~ 0.018 & 19.49~ 0.27 & 1.43 & 1.507 & 19.513 & 1.457 & 19.384 &  2.243~ 0.047 & 0.304~ 0.012 \\
E322-008 & 3029 &$-$2 & 0.363 & 1.205~ 0.023 & 20.81~ 0.54 & 2.53 & 1.395 & 20.086 & 1.513 & 20.556 &  \dots~ \dots & \dots~ \dots \\
E322-038 & 3134 &$-$3 & 0.560 & 1.242~ 0.051 & 20.32~ 0.32 & 2.37 & 1.471 & 20.765 & 1.297 & 20.101 &  2.311~ 0.041 & 0.315~ 0.014 \\
E322-059 & 3240 &$-$2 & 0.796 & 1.220~ 0.027 & 20.57~ 0.28 & 3.64 & 1.600 & 20.637 & 1.458 & 20.188 &  2.363~ 0.041 & 0.311~ 0.014 \\
E322-088 & 2606 &$-$2 & 0.484 & 1.156~ 0.034 & 19.85~ 0.68 & 2.46 & 1.021 & 18.865 & 0.963 & 18.815 &  2.208~ 0.041 & 0.278~ 0.014 \\
E322-101 & 2053 &  0 & 0.493 & 1.167~ 0.023 & \dots~ \dots& 1.99 & \dots &  \dots & 1.256 & 20.453 &  2.207~ 0.021 & 0.305~ 0.008 \\
E323-034 & 4254 &$-$5 & 0.535 & 1.165~ 0.019 & 20.86~ 0.28 & 2.51 & 1.335 & 19.351 & 1.386 & 19.504 &  2.375~ 0.013 & 0.293~ 0.006 \\
E358-006 & 1358 &$-$4 & 0.043 & 1.068~ 0.020 & 17.59~ 0.37 & 0.97 & 1.494 & 21.631 & 1.041 & 20.076 &  \dots~ \dots & \dots~ \dots \\
E358-059 & 1030 &$-$3 & 0.039 & 1.081~ 0.016 & 17.27~ 0.27 & 1.02 & 1.067 & 20.145 & 1.008 & 19.948 &  \dots~ \dots & 0.140~ 0.008 \\
I0745    & 1135 &$-$2 & 0.091 & 0.978~ 0.010 & 16.63~ 0.36 & 2.33 & 1.117 & 20.222 & 1.205 & 20.443 &  \dots~ \dots & \dots~ \dots \\
I1153    &  744 &$-$2 & 0.083 & 1.197~ 0.011 & 18.41~ 0.28 & 0.83 & 1.268 & 20.276 & 1.117 & 19.752 &  \dots~ \dots & \dots~ \dots \\
I1459    & 1624 &$-$5 & 0.069 & 1.194~ 0.018 & 22.21~ 0.33 & 1.49 & 1.684 & 19.574 & 1.718 & 19.752 &  2.492~ 0.042 & 0.332~ 0.010 \\
I2006    & 1340 &$-$5 & 0.048 & 1.183~ 0.018 & 19.61~ 0.34 & 1.45 & 1.259 & 19.246 & 1.274 & 19.300 &  2.086~ 0.013 & 0.273~ 0.004 \\
I2311    & 1836 &$-$5 & 0.620 & 1.136~ 0.027 & 19.98~ 0.24 & 2.09 & 1.275 & 18.975 & 1.238 & 18.896 &  \dots~ \dots & \dots~ \dots \\
I3370    & 2971 &$-$5 & 0.404 & 1.185~ 0.022 & 20.91~ 0.24 & 2.69 & 1.425 & 19.332 & 1.483 & 19.560 &  2.318~ 0.019 & 0.257~ 0.008 \\
I4296    & 3660 &$-$5 & 0.266 & 1.236~ 0.034 & 22.64~ 0.27 & 3.15 & 1.683 & 20.154 & 1.700 & 20.203 &  2.532~ 0.013 & 0.330~ 0.005 \\
I4797    & 2597 &$-$4 & 0.340 & 1.184~ 0.018 & 20.59~ 0.27 & 1.96 & 1.331 & 19.282 & 1.382 & 19.266 &  2.366~ 0.041 & 0.302~ 0.014 \\
I4889    & 2486 &$-$5 & 0.228 & 1.137~ 0.016 & 20.64~ 0.24 & 1.77 & 1.356 & 19.207 & 1.347 & 19.173 &  2.219~ 0.015 & 0.254~ 0.004 \\
I4943    & 2930 &$-$5 & 0.221 & 1.133~ 0.023 & 19.23~ 0.25 & 2.80 & 1.153 & 19.727 & 1.000 & 19.209 &  2.230~ 0.042 & 0.256~ 0.010 \\
I5328    & 3136 &$-$5 & 0.062 & 1.173~ 0.023 & 21.16~ 0.22 & 2.67 & 1.411 & 19.695 & 1.446 & 19.818 &  2.303~ 0.042 & 0.294~ 0.010 \\
N0063    & 1168 &  0 & 0.481 & 0.979~ 0.018 & 18.25~ 0.37 & 1.14 & 1.380 & 19.974 & 1.919 & 21.554 &  \dots~ \dots & \dots~ \dots \\
N0185    &$-$227 &$-$5 & 0.792 & 1.051~ 0.017 & 14.17~ 0.24 & 0.01 & 2.129 & 20.828 & \dots &  \dots &  \dots~ \dots & \dots~ \dots \\
N0221    &$-$200 &$-$6 & 0.346 & 1.133~ 0.007 & 15.60~ 0.21 & 0.01 & 1.475 & 17.079 & 1.453 & 16.893 &  1.828~ 0.009 & 0.119~ 0.003 \\
N0274    & 1729 &$-$3 & 0.242 & 1.135~ 0.020 & 18.70~ 0.49 & 1.68 & 1.228 & 19.624 & 1.519 & 20.631 &  \dots~ \dots & \dots~ \dots \\
N0404    & $-$36 &$-$3 & 0.253 & 1.054~ 0.011 & 16.63~ 0.21 & 0.01 & 1.711 & 19.858 & 1.575 & 19.424 &  1.676~ 0.039 & 0.016~ 0.007 \\
N0448    & 1917 &$-$3 & 0.261 & 1.132~ 0.029 & \dots~ \dots& 1.41 & \dots &  \dots & 1.163 & 19.316 &  \dots~ \dots & \dots~ \dots \\
N0524    & 2416 &$-$1 & 0.357 & 1.221~ 0.010 & 21.56~ 0.28 & 1.65 & 1.551 & 19.252 & 1.579 & 19.365 &  \dots~ \dots & \dots~ \dots \\
N0584    & 1875 &$-$5 & 0.182 & 1.157~ 0.009 & 20.62~ 0.28 & 1.41 & 1.433 & 18.915 & 1.445 & 18.970 &  2.295~ 0.009 & 0.268~ 0.003 \\
N0596    & 1817 &$-$4 & 0.159 & 1.135~ 0.008 & 20.27~ 0.22 & 1.51 & 1.526 & 19.789 & 1.509 & 19.737 &  2.179~ 0.024 & 0.237~ 0.006 \\
N0636    & 1805 &$-$5 & 0.110 & 1.156~ 0.008 & 20.64~ 0.24 & 1.13 & 1.499 & 20.067 & 1.495 & 20.080 &  2.212~ 0.011 & 0.259~ 0.006 \\
N0661    & 3836 &$-$4 & 0.296 & 1.164~ 0.011 & 19.60~ 0.31 & 2.73 & 1.182 & 19.340 & 1.258 & 19.583 &  2.250~ 0.026 & 0.293~ 0.008 \\
N0680    & 2953 &$-$4 & 0.331 & 1.189~ 0.011 & 20.48~ 0.45 & 2.87 & 1.269 & 19.306 & 1.378 & 19.706 &  2.292~ 0.026 & 0.281~ 0.008 \\
N0720    & 1716 &$-$5 & 0.070 & 1.214~ 0.009 & 21.66~ 0.25 & 1.09 & 1.544 & 19.402 & 1.523 & 19.333 &  2.392~ 0.009 & 0.324~ 0.004 \\
N0821    & 1716 &$-$5 & 0.475 & 1.196~ 0.022 & 20.88~ 0.24 & 1.00 & 1.510 & 19.627 & 1.594 & 19.915 &  2.288~ 0.009 & 0.294~ 0.004 \\
N0855    &  600 &$-$5 & 0.311 & 1.015~ 0.018 & 16.25~ 0.24 & 0.23 & 1.365 & 20.660 & 1.246 & 20.344 &  \dots~ \dots & \dots~ \dots \\
N0891    &  529 &  3 & 0.280 & 1.142~ 0.017 & 19.23~ 0.23 & 0.36 & 2.081 & 21.577 & 2.110 & 21.713 &  \dots~ \dots & \dots~ \dots \\
N0936    & 1434 &$-$1 & 0.152 & 1.213~ 0.010 & 21.78~ 0.34 & 0.94 & 1.838 & 20.357 & 1.948 & 20.663 &  2.263~ 0.014 & 0.276~ 0.004 \\
N0949    &  610 &  3 & 0.254 & 0.958~ 0.011 & 16.96~ 0.26 & 0.44 & 1.374 & 20.026 & 1.611 & 20.755 &  \dots~ \dots & \dots~ \dots \\
N1023    &  648 &$-$3 & 0.263 & 1.193~ 0.017 & 20.70~ 0.24 & 0.25 & 1.706 & 19.161 & 1.667 & 19.011 &  2.259~ 0.030 & 0.298~ 0.009 \\
N1052    & 1475 &$-$5 & 0.115 & 1.213~ 0.010 & 20.63~ 0.33 & 1.48 & 1.431 & 19.090 & 1.390 & 18.974 &  2.302~ 0.032 & 0.300~ 0.007 \\
N1162    & 2331 &$-$5 & 0.206 & 1.173~ 0.032 & 20.63~ 0.34 & 1.91 & 1.245 & 19.585 & 1.294 & 19.758 &  \dots~ \dots & \dots~ \dots \\
N1172    & 1669 &$-$4 & 0.275 & 1.112~ 0.032 & 19.40~ 0.24 & 1.40 & 1.664 & 21.203 & 1.991 & 22.321 &  2.073~ 0.039 & 0.230~ 0.007 \\
N1199    & 2705 &$-$5 & 0.234 & 1.188~ 0.012 & 20.93~ 0.38 & 2.03 & 1.420 & 19.753 & 1.400 & 19.728 &  2.313~ 0.039 & 0.293~ 0.007 \\
N1201    & 1720 &$-$2 & 0.070 & 1.178~ 0.009 & 20.54~ 0.35 & 2.04 & 1.557 & 19.725 & 1.503 & 19.546 &  \dots~ \dots & \dots~ \dots \\
N1209    & 2619 &$-$5 & 0.163 & 1.198~ 0.010 & 21.01~ 0.30 & 1.72 & 1.253 & 19.050 & 1.187 & 18.829 &  2.321~ 0.055 & 0.301~ 0.010 \\
N1297    & 1550 &$-$2 & 0.119 & 1.187~ 0.018 & 20.25~ 0.46 & 1.42 & 1.578 & 20.915 & 1.853 & 21.832 &  \dots~ \dots & \dots~ \dots \\
N1316    & 1780 &$-$2 & 0.090 & 1.132~ 0.016 & 22.65~ 0.25 & 1.79 & 1.964 & 19.541 & 1.986 & 19.663 &  \dots~ \dots & \dots~ \dots \\
N1332    & 1469 &$-$3 & 0.137 & 1.222~ 0.010 & 21.37~ 0.26 & 1.01 & 1.450 & 18.851 & 1.477 & 18.950 &  \dots~ \dots & \dots~ \dots \\
N1336    & 1466 &$-$3 & 0.049 & 1.124~ 0.032 & 18.46~ 0.25 & 1.45 & 1.606 & 21.632 & 1.533 & 21.395 &  \dots~ \dots & \dots~ \dots \\
N1339    & 1380 &$-$4 & 0.057 & 1.134~ 0.012 & 19.07~ 0.40 & 0.95 & 1.189 & 19.220 & 1.278 & 19.523 &  2.208~ 0.012 & 0.294~ 0.003 \\
N1344    & 1217 &$-$5 & 0.078 & 1.135~ 0.011 & 20.52~ 0.35 & 0.97 & 1.521 & 19.308 & 1.474 & 19.130 &  \dots~ \dots & \dots~ \dots \\
N1351    & 1539 &$-$3 & 0.061 & 1.148~ 0.016 & 19.76~ 0.24 & 1.12 & 1.539 & 20.343 & 1.589 & 20.516 &  2.156~ 0.014 & 0.262~ 0.004 \\
N1366    & 1310 &$-$2 & 0.073 & 1.095~ 0.018 & 18.93~ 0.34 & 1.48 & 1.230 & 19.383 & 1.316 & 19.745 &  \dots~ \dots & \dots~ \dots \\
N1373    & 1385 &$-$4 & 0.061 & 1.085~ 0.013 & 17.48~ 0.50 & 2.05 & 1.004 & 19.797 & 1.028 & 19.863 &  \dots~ \dots & \dots~ \dots \\
N1374    & 1397 &$-$5 & 0.061 & 1.146~ 0.016 & 19.77~ 0.22 & 1.00 & 1.375 & 19.384 & 1.404 & 19.533 &  2.265~ 0.014 & 0.304~ 0.004 \\
N1375    &  773 &$-$2 & 0.062 & 1.070~ 0.019 & 17.93~ 0.22 & 0.51 & 1.174 & 19.943 & 1.156 & 19.960 &  \dots~ \dots & \dots~ \dots \\
N1379    & 1408 &$-$5 & 0.052 & 1.143~ 0.019 & 19.89~ 0.23 & 1.00 & 1.465 & 19.722 & 1.476 & 19.807 &  2.094~ 0.015 & 0.248~ 0.004 \\
N1381    & 1751 &$-$2 & 0.059 & 1.189~ 0.018 & 19.44~ 0.28 & 1.16 & 1.200 & 18.842 & 1.238 & 18.960 &  \dots~ \dots & \dots~ \dots \\
N1382    & 1763 &$-$3 & 0.074 & 1.106~ 0.013 & 17.80~ 0.36 & 2.58 & 1.182 & 20.471 & 1.196 & 20.202 &  \dots~ \dots & \dots~ \dots \\
N1387    & 1302 &$-$3 & 0.056 & 1.208~ 0.047 & 20.79~ 0.24 & 1.16 & 1.551 & 19.618 & 1.747 & 20.372 &  \dots~ \dots & \dots~ \dots \\
N1389    &  950 &$-$3 & 0.045 & 1.145~ 0.019 & 19.60~ 0.26 & 0.91 & 1.166 & 18.718 & \dots &  \dots &  \dots~ \dots & \dots~ \dots \\
N1395    & 1702 &$-$5 & 0.100 & 1.215~ 0.010 & 21.83~ 0.24 & 1.49 & 1.690 & 19.676 & 1.647 & 19.520 &  2.396~ 0.014 & 0.312~ 0.003 \\
\noalign{\vspace{3pt}\hrule\vspace{3pt}}
\end{tabular}\end{minipage}\end{table*}
\newpage
\begin{table*}\setcounter{table}{0}
\centering \begin{minipage}{172mm}
\caption{Fundamental Plane Data for SBF Survey Galaxies\,---\,\rlap{\it continued}}
\tabcolsep=0.15cm\footnotesize
\begin{tabular}{lrrcccccccccc}
\noalign{\vspace{3pt}\hrule\vspace{3pt}}
Galaxy & $v_h\;$ & $T$ & $A_B$ & \viz~ $\pm$~ & ~\Nbar~~~ $\pm$ & \PD\ &
 $\log R_e^{(I)}$ & $\langle \mu_R \rangle_e^{(I)}$ & $\log R_e^{(V)}$ & $\langle \mu_R \rangle_e^{(V)}$ &
 $\log\sigma$~~ $\pm$~~ & Mg$_2$~~~ $\pm$~ \\
\noalign{\vspace{3pt}\hrule\vspace{3pt}}
N1399    & 1422 &$-$5 & 0.056 & 1.227~ 0.016 & 21.69~ 0.24 & 1.34 & 1.581 & 18.920 & 1.619 & 19.061 &  2.509~ 0.013 & 0.329~ 0.003 \\
N1400    &  549 &$-$3 & 0.279 & 1.170~ 0.009 & 20.93~ 0.38 & 0.32 & 1.470 & 19.461 & 1.500 & 19.641 &  2.385~ 0.018 & 0.287~ 0.004 \\
N1404    & 1925 &$-$5 & 0.049 & 1.224~ 0.016 & 21.54~ 0.26 & 2.00 & 1.440 & 18.449 & 1.480 & 18.650 &  2.366~ 0.012 & 0.304~ 0.003 \\
N1407    & 1766 &$-$5 & 0.298 & 1.222~ 0.010 & 22.44~ 0.32 & 1.42 & 1.761 & 19.837 & 1.722 & 19.725 &  2.464~ 0.019 & 0.328~ 0.005 \\
N1419    & 1557 &$-$5 & 0.056 & 1.110~ 0.018 & 17.81~ 0.30 & 1.63 & 1.056 & 19.503 & 0.988 & 19.260 &  2.141~ 0.029 & 0.242~ 0.006 \\
N1426    & 1443 &$-$5 & 0.072 & 1.161~ 0.009 & 20.10~ 0.26 & 1.27 & 1.485 & 20.104 & 1.476 & 20.090 &  2.172~ 0.025 & 0.258~ 0.006 \\
N1427    & 1430 &$-$4 & 0.051 & 1.152~ 0.018 & 20.50~ 0.30 & 1.45 & 1.598 & 20.174 & 1.549 & 20.006 &  2.197~ 0.018 & 0.244~ 0.005 \\
N1439    & 1670 &$-$5 & 0.129 & 1.131~ 0.009 & 20.38~ 0.24 & 1.25 & 1.674 & 20.840 & 1.893 & 21.605 &  2.140~ 0.022 & 0.271~ 0.005 \\
N1521    & 4165 &$-$5 & 0.176 & 1.152~ 0.029 & 22.12~ 0.41 & 2.81 & 1.515 & 20.378 & 1.523 & 20.397 &  2.354~ 0.029 & 0.274~ 0.006 \\
N1527    & 1165 &$-$3 & 0.054 & 1.230~ 0.016 & 20.50~ 0.28 & 0.95 & 1.474 & 19.378 & 1.493 & 19.424 &  2.211~ 0.025 & 0.262~ 0.007 \\
N1537    & 1371 &$-$3 & 0.107 & 1.096~ 0.018 & 20.45~ 0.26 & 1.72 & 1.547 & 19.652 & \dots &  \dots &  2.198~ 0.029 & 0.271~ 0.006 \\
N1543    & 1088 &$-$2 & 0.117 & 1.173~ 0.016 & 20.92~ 0.24 & 1.11 & 1.711 & 20.086 & 1.685 & 20.029 &  2.181~ 0.025 & 0.280~ 0.007 \\
N1549    & 1153 &$-$5 & 0.055 & 1.168~ 0.016 & 21.54~ 0.25 & 1.13 & 1.752 & 19.601 & 1.815 & 19.814 &  2.320~ 0.019 & \dots~ \dots \\
N1553    & 1280 &$-$2 & 0.053 & 1.159~ 0.016 & 21.47~ 0.24 & 1.31 & 1.533 & 18.403 & 1.555 & 18.480 &  2.230~ 0.036 & 0.268~ 0.010 \\
N1574    & 1042 &$-$3 & 0.070 & 1.162~ 0.016 & 20.87~ 0.28 & 1.10 & 1.590 & 19.455 & 1.717 & 19.884 &  2.329~ 0.036 & 0.290~ 0.010 \\
N1587    & 3667 &$-$5 & 0.308 & 1.215~ 0.012 & 21.43~ 0.87 & 3.33 & 1.339 & 19.578 & 1.323 & 19.502 &  2.341~ 0.039 & 0.315~ 0.007 \\
N1596    & 1523 &$-$2 & 0.041 & 1.171~ 0.016 & 19.33~ 0.24 & 1.57 & 1.211 & 18.563 & 1.209 & 18.599 &  2.212~ 0.036 & 0.275~ 0.010 \\
N1653    & 4343 &$-$4 & 0.178 & 1.167~ 0.011 & 21.37~ 0.32 & 3.00 & 1.407 & 20.151 & 1.408 & 20.157 &  2.342~ 0.055 & 0.284~ 0.010 \\
N1700    & 3881 &$-$5 & 0.185 & 1.163~ 0.011 & 21.57~ 0.24 & 2.88 & 1.344 & 19.213 & 1.301 & 19.027 &  2.372~ 0.009 & 0.279~ 0.004 \\
N2271    & 2588 &$-$3 & 0.527 & 1.234~ 0.029 & 20.64~ 0.32 & 1.93 & 1.241 & 19.300 & 1.246 & 19.296 &  2.194~ 0.047 & 0.284~ 0.012 \\
N2293    & 1994 &$-$1 & 0.518 & 1.237~ 0.013 & 20.44~ 0.45 & 2.03 & 1.664 & 20.284 & 1.488 & 19.714 &  2.424~ 0.020 & 0.309~ 0.005 \\
N2305    & 3571 &$-$5 & 0.327 & 1.225~ 0.027 & 20.61~ 0.27 & 3.60 & 1.561 & 20.254 & 1.473 & 19.958 &  2.361~ 0.047 & 0.299~ 0.012 \\
N2325    & 2248 &$-$5 & 0.508 & 1.164~ 0.014 & 21.03~ 0.24 & 2.14 & 1.747 & 20.609 & 1.568 & 20.059 &  2.143~ 0.036 & 0.287~ 0.008 \\
N2380    & 1775 &$-$2 & 1.318 & 1.110~ 0.009 & 21.17~ 0.40 & 1.63 & 1.613 & 19.567 & 1.537 & 19.313 &  2.291~ 0.029 & 0.271~ 0.008 \\
N2434    & 1327 &$-$5 & 1.073 & 1.098~ 0.055 & 20.53~ 0.25 & 1.43 & 1.633 & 19.868 & 1.565 & 19.674 &  2.321~ 0.022 & 0.247~ 0.008 \\
N2549    & 1056 &$-$2 & 0.282 & 1.163~ 0.012 & 18.85~ 0.34 & 1.39 & 1.229 & 18.687 & 1.238 & 18.663 &  \dots~ \dots & \dots~ \dots \\
N2592    & 1989 &$-$5 & 0.261 & 1.205~ 0.010 & 19.37~ 0.47 & 2.26 & 1.028 & 18.893 & 1.122 & 19.147 &  \dots~ \dots & \dots~ \dots \\
N2634    & 2269 &$-$5 & 0.094 & 1.154~ 0.022 & 20.50~ 0.70 & 2.22 & 1.636 & 21.126 & 1.605 & 21.056 &  2.250~ 0.017 & 0.279~ 0.004 \\
N2681    &  715 &  0 & 0.098 & 1.041~ 0.010 & 20.27~ 0.39 & 0.74 & 1.949 & 20.926 & 1.892 & 20.718 &  \dots~ \dots & \dots~ \dots \\
N2695    & 1825 &$-$2 & 0.077 & 1.183~ 0.051 & 20.32~ 0.38 & 1.64 & 1.279 & 19.599 & 1.312 & 19.692 &  2.278~ 0.055 & 0.311~ 0.010 \\
N2699    & 1825 &$-$5 & 0.087 & 1.152~ 0.051 & 19.27~ 0.24 & 1.68 & 1.030 & 18.855 & 0.962 & 18.705 &  2.114~ 0.055 & 0.266~ 0.010 \\
N2768    & 1363 &$-$5 & 0.192 & 1.144~ 0.027 & 21.04~ 0.28 & 0.96 & 1.604 & 19.515 & 1.589 & 19.561 &  2.252~ 0.021 & 0.262~ 0.005 \\
N2784    &  708 &$-$2 & 0.927 & 1.188~ 0.023 & 20.29~ 0.30 & 1.37 & 1.608 & 18.743 & 1.671 & 18.961 &  \dots~ \dots & \dots~ \dots \\
N2787    &  689 &$-$1 & 0.567 & 1.194~ 0.019 & 18.75~ 0.40 & 1.18 & 1.392 & 18.622 & 1.520 & 19.111 &  \dots~ \dots & \dots~ \dots \\
N2865    & 2581 &$-$5 & 0.356 & 1.105~ 0.019 & 20.82~ 0.26 & 2.14 & 1.439 & 19.838 & 1.488 & 19.954 &  2.246~ 0.008 & 0.188~ 0.003 \\
N2904    & 2387 &$-$3 & 0.544 & 1.201~ 0.041 & 19.28~ 0.24 & 2.07 & 1.203 & 19.658 & 1.146 & 19.483 &  \dots~ \dots & \dots~ \dots \\
N2950    & 1327 &$-$2 & 0.072 & 1.110~ 0.019 & 19.32~ 0.32 & 2.23 & 1.360 & 18.968 & 1.378 & 18.988 &  \dots~ \dots & \dots~ \dots \\
N2962    & 2117 &$-$1 & 0.249 & 1.176~ 0.017 & 20.32~ 0.38 & 2.06 & 1.569 & 20.647 & 1.700 & 21.174 &  \dots~ \dots & \dots~ \dots \\
N2974    & 1924 &$-$5 & 0.235 & 1.203~ 0.015 & 20.96~ 0.30 & 2.15 & 1.656 & 20.060 & 1.575 & 19.774 &  2.366~ 0.023 & 0.290~ 0.004 \\
N3031    & $-$38 &  2 & 0.347 & 1.187~ 0.011 & 19.07~ 0.32 & 0.04 & 1.405 & 16.930 & 1.412 & 17.041 &  \dots~ \dots & \dots~ \dots \\
N3032    & 1561 &$-$2 & 0.073 & 1.073~ 0.019 & 18.78~ 0.33 & 1.71 & 1.525 & 20.979 & 1.311 & 20.195 &  \dots~ \dots & \dots~ \dots \\
N3056    & 1047 &$-$1 & 0.386 & 1.073~ 0.023 & 17.90~ 0.30 & 1.98 & 1.310 & 19.510 & 1.613 & 20.391 &  \dots~ \dots & \dots~ \dots \\
N3073    & 1154 &$-$3 & 0.044 & 1.007~ 0.019 & 18.29~ 0.94 & 1.51 & 1.417 & 21.538 & 1.362 & 21.291 &  \dots~ \dots & \dots~ \dots \\
N3078    & 2506 &$-$5 & 0.307 & 1.209~ 0.017 & 21.66~ 0.35 & 2.67 & 1.522 & 19.769 & 1.563 & 19.900 &  2.399~ 0.019 & 0.321~ 0.005 \\
N3087    & 2662 &$-$4 & 0.454 & 1.164~ 0.019 & 21.32~ 0.27 & 2.23 & 1.501 & 19.770 & 1.360 & 19.281 &  2.433~ 0.055 & 0.279~ 0.010 \\
N3115    &  698 &$-$3 & 0.205 & 1.183~ 0.010 & 20.46~ 0.21 & 0.95 & 1.590 & 18.418 & 1.595 & 18.379 &  2.394~ 0.009 & 0.289~ 0.003 \\
N3136    & 1647 &$-$5 & 1.027 & 1.095~ 0.033 & 21.24~ 0.25 & 1.39 & 1.722 & 19.848 & 1.700 & 19.751 &  2.376~ 0.022 & 0.262~ 0.008 \\
N3156    & 1296 &$-$2 & 0.148 & 1.011~ 0.011 & 18.28~ 0.24 & 1.34 & 1.274 & 19.950 & 1.193 & 19.732 &  2.035~ 0.039 & 0.080~ 0.007 \\
N3193    & 1378 &$-$5 & 0.111 & 1.174~ 0.009 & 21.36~ 0.26 & 0.85 & 1.472 & 19.603 & 1.453 & 19.506 &  2.302~ 0.055 & 0.282~ 0.010 \\
N3226    & 1275 &$-$5 & 0.098 & 1.178~ 0.015 & 20.59~ 0.30 & 1.33 & 1.792 & 21.199 & \dots &  \dots &  2.296~ 0.039 & 0.294~ 0.007 \\
N3245    & 1358 &$-$2 & 0.108 & 1.139~ 0.023 & 20.03~ 0.26 & 1.79 & 1.302 & 18.844 & 1.381 & 19.129 &  \dots~ \dots & \dots~ \dots \\
N3250    & 2883 &$-$5 & 0.445 & 1.226~ 0.019 & 22.26~ 0.24 & 2.14 & 1.451 & 19.322 & 1.418 & 19.212 &  2.431~ 0.018 & 0.308~ 0.006 \\
N3257    & 3023 &$-$3 & 0.334 & 1.192~ 0.041 & 19.77~ 0.27 & 2.61 & 1.100 & 19.409 & 1.060 & 19.304 &  \dots~ \dots & \dots~ \dots \\
N3258    & 2808 &$-$5 & 0.362 & 1.209~ 0.034 & 21.20~ 0.30 & 2.32 & 1.603 & 20.439 & 1.565 & 20.309 &  2.457~ 0.017 & 0.337~ 0.005 \\
N3268    & 2761 &$-$5 & 0.446 & 1.189~ 0.023 & 21.29~ 0.30 & 1.85 & 1.638 & 20.585 & 1.608 & 20.517 &  2.383~ 0.033 & 0.318~ 0.009 \\
N3309    & 4068 &$-$5 & 0.344 & 1.208~ 0.021 & 21.79~ 0.70 & 3.40 & 1.335 & 19.613 & 1.375 & 19.754 &  2.395~ 0.013 & 0.325~ 0.005 \\
N3311    & 3713 &$-$4 & 0.343 & 1.229~ 0.034 & 22.76~ 0.53 & 3.12 & 2.098 & 22.138 & 2.003 & 21.861 &  2.272~ 0.021 & 0.333~ 0.010 \\
N3368    &  899 &  2 & 0.109 & 1.145~ 0.015 & 20.23~ 0.28 & 1.63 & 1.698 & 19.140 & 1.713 & 19.241 &  \dots~ \dots & \dots~ \dots \\
N3377    &  689 &$-$5 & 0.148 & 1.114~ 0.009 & 19.24~ 0.21 & 0.89 & 1.610 & 19.715 & 1.618 & 19.748 &  2.105~ 0.006 & 0.238~ 0.002 \\
N3379    &  922 &$-$5 & 0.105 & 1.193~ 0.015 & 20.48~ 0.21 & 1.59 & 1.708 & 19.217 & 1.657 & 18.973 &  2.308~ 0.005 & 0.289~ 0.002 \\
N3384    &  728 &$-$3 & 0.114 & 1.151~ 0.018 & 20.13~ 0.22 & 1.24 & 1.755 & 19.802 & 1.645 & 19.467 &  2.149~ 0.009 & 0.266~ 0.003 \\
N3412    &  867 &$-$2 & 0.121 & 1.111~ 0.015 & 18.74~ 0.23 & 1.40 & 1.375 & 19.035 & 1.700 & 20.063 &  1.995~ 0.011 & 0.207~ 0.004 \\
N3414    & 1476 &$-$2 & 0.104 & 1.149~ 0.019 & 20.63~ 0.38 & 1.66 & 1.546 & 19.927 & 1.619 & 20.200 &  \dots~ \dots & \dots~ \dots \\
N3457    & 1156 &  0 & 0.135 & 1.098~ 0.015 & \dots~ \dots& 1.32 & \dots &  \dots & 1.036 & 18.861 &  \dots~ \dots & \dots~ \dots \\
\noalign{\vspace{3pt}\hrule\vspace{3pt}}
\end{tabular}\end{minipage}\end{table*}
\newpage
\begin{table*}\setcounter{table}{0}
\centering \begin{minipage}{172mm}
\caption{Fundamental Plane Data for SBF Survey Galaxies\,---\,\rlap{\it continued}}
\tabcolsep=0.15cm\footnotesize
\begin{tabular}{lrrcccccccccc}
\noalign{\vspace{3pt}\hrule\vspace{3pt}}
Galaxy & $v_h\;$ & $T$ & $A_B$ & \viz~ $\pm$~ & ~\Nbar~~~ $\pm$ & \PD\ &
 $\log R_e^{(I)}$ & $\langle \mu_R \rangle_e^{(I)}$ & $\log R_e^{(V)}$ & $\langle \mu_R \rangle_e^{(V)}$ &
 $\log\sigma$~~ $\pm$~~ & Mg$_2$~~~ $\pm$~ \\
\noalign{\vspace{3pt}\hrule\vspace{3pt}}
N3489    &  693 &$-$1 & 0.071 & 1.041~ 0.023 & 18.87~ 0.22 & 0.96 & 1.287 & 18.252 & 1.296 & 18.222 &  2.007~ 0.010 & 0.153~ 0.004 \\
N3585    & 1491 &$-$5 & 0.278 & 1.160~ 0.016 & 21.46~ 0.26 & 2.31 & 1.755 & 19.685 & 1.741 & 19.640 &  2.317~ 0.026 & 0.297~ 0.007 \\
N3599    &  850 &$-$2 & 0.089 & 1.112~ 0.012 & 19.10~ 0.26 & 1.14 & 1.620 & 21.183 & 1.720 & 21.574 &  1.883~ 0.039 & 0.159~ 0.007 \\
N3605    &  686 &$-$5 & 0.089 & 1.118~ 0.024 & 18.12~ 0.34 & 1.11 & 0.997 & 19.116 & 0.987 & 19.052 &  2.049~ 0.039 & 0.184~ 0.007 \\
N3607    &  951 &$-$2 & 0.090 & 1.152~ 0.010 & 21.35~ 0.25 & 1.40 & 1.620 & 19.378 & 1.607 & 19.400 &  2.379~ 0.039 & 0.285~ 0.007 \\
N3608    & 1197 &$-$5 & 0.090 & 1.156~ 0.009 & 20.51~ 0.23 & 0.92 & 1.546 & 19.875 & 1.604 & 20.094 &  2.269~ 0.009 & 0.295~ 0.004 \\
N3610    & 1765 &$-$5 & 0.043 & 1.108~ 0.015 & 19.91~ 0.28 & 2.21 & 1.329 & 18.997 & 1.302 & 18.893 &  2.192~ 0.032 & 0.245~ 0.006 \\
N3613    & 2054 &$-$5 & 0.053 & 1.175~ 0.015 & 20.97~ 0.44 & 2.22 & 1.502 & 19.788 & 1.363 & 19.279 &  2.319~ 0.039 & 0.271~ 0.007 \\
N3617    & 2224 &$-$4 & 0.213 & 1.050~ 0.023 & 19.15~ 0.32 & 3.07 & 1.465 & 21.106 & 1.464 & 21.107 &  \dots~ \dots & \dots~ \dots \\
N3640    & 1302 &$-$5 & 0.191 & 1.140~ 0.009 & 21.28~ 0.23 & 1.49 & 1.579 & 19.531 & 1.553 & 19.455 &  2.236~ 0.039 & 0.247~ 0.007 \\
N3641    & 1758 &$-$5 & 0.182 & 1.131~ 0.012 & 18.62~ 0.31 & 1.90 & 1.407 & 21.258 & 1.240 & 20.653 &  2.207~ 0.055 & 0.284~ 0.010 \\
N3818    & 1498 &$-$5 & 0.155 & 1.124~ 0.015 & 20.32~ 0.63 & 1.69 & 1.346 & 19.886 & 1.434 & 20.142 &  2.248~ 0.010 & 0.305~ 0.005 \\
N3904    & 1750 &$-$5 & 0.311 & 1.156~ 0.055 & 20.96~ 0.24 & 1.51 & 1.374 & 19.014 & 1.381 & 19.037 &  2.354~ 0.021 & 0.293~ 0.006 \\
N3923    & 1607 &$-$5 & 0.358 & 1.194~ 0.055 & 21.91~ 0.23 & 1.50 & 1.690 & 19.382 & 1.706 & 19.429 &  2.334~ 0.021 & 0.297~ 0.006 \\
N3928    &  974 &  3 & 0.085 & 1.096~ 0.015 & 17.95~ 0.66 & 1.47 & 1.107 & 19.285 & 1.211 & 19.631 &  \dots~ \dots & \dots~ \dots \\
N3941    &  944 &$-$2 & 0.091 & 1.125~ 0.013 & 19.28~ 0.26 & 1.63 & 1.322 & 18.464 & 1.366 & 18.606 &  \dots~ \dots & \dots~ \dots \\
N3962    & 1822 &$-$5 & 0.201 & 1.145~ 0.023 & 21.96~ 0.52 & 2.85 & 1.795 & 20.545 & 1.683 & 20.190 &  2.317~ 0.039 & 0.298~ 0.007 \\
N3990    &  705 &$-$3 & 0.069 & 1.151~ 0.019 & \dots~ \dots& 0.98 & \dots &  \dots & 0.967 & 18.855 &  \dots~ \dots & \dots~ \dots \\
N3998    & 1028 &$-$2 & 0.069 & 1.194~ 0.011 & 19.84~ 0.27 & 1.35 & 1.353 & 18.712 & 1.458 & 19.102 &  \dots~ \dots & \dots~ \dots \\
N4024    & 1694 &$-$3 & 0.182 & 1.141~ 0.016 & 20.00~ 0.51 & 3.44 & 1.185 & 19.098 & 1.452 & 20.043 &  2.165~ 0.039 & 0.241~ 0.007 \\
N4033    & 1521 &$-$5 & 0.205 & 1.113~ 0.023 & 18.98~ 0.28 & 2.56 & 1.266 & 19.593 & 1.326 & 19.652 &  2.070~ 0.039 & 0.236~ 0.007 \\
N4105    & 1882 &$-$5 & 0.263 & 1.171~ 0.017 & 21.25~ 0.24 & 1.89 & 1.627 & 19.914 & 1.573 & 19.699 &  2.398~ 0.028 & 0.292~ 0.006 \\
N4111    &  806 &$-$1 & 0.063 & 1.096~ 0.015 & 19.07~ 0.30 & 1.20 & 1.267 & 18.709 & 1.155 & 17.923 &  \dots~ \dots & \dots~ \dots \\
N4125    & 1340 &$-$5 & 0.082 & 1.174~ 0.011 & 21.80~ 0.31 & 1.86 & 1.772 & 19.890 & 1.875 & 20.241 &  2.347~ 0.055 & 0.281~ 0.010 \\
N4138    &  835 &$-$1 & 0.060 & 1.164~ 0.013 & 18.95~ 0.32 & 1.01 & 1.272 & 19.000 & 1.294 & 19.114 &  \dots~ \dots & \dots~ \dots \\
N4150    &  244 &$-$2 & 0.078 & 1.071~ 0.017 & 18.06~ 0.30 & 0.81 & 1.189 & 19.020 & 1.290 & 19.400 &  \dots~ \dots & \dots~ \dots \\
N4203    & 1117 &$-$3 & 0.052 & 1.195~ 0.019 & 20.73~ 0.25 & 1.26 & 1.945 & 20.940 & 1.953 & 20.950 &  \dots~ \dots & \dots~ \dots \\
N4233    & 2371 &$-$2 & 0.103 & 1.191~ 0.015 & 20.30~ 0.95 & 2.43 & 1.203 & 19.364 & 1.342 & 19.891 &  \dots~ \dots & \dots~ \dots \\
N4261    & 2200 &$-$5 & 0.078 & 1.258~ 0.014 & 22.06~ 0.27 & 1.28 & 1.505 & 19.321 & 1.380 & 18.892 &  2.483~ 0.010 & 0.326~ 0.005 \\
N4278    &  643 &$-$5 & 0.123 & 1.161~ 0.012 & 20.39~ 0.27 & 0.66 & 1.477 & 18.912 & 1.528 & 19.141 &  2.376~ 0.010 & 0.276~ 0.005 \\
N4283    & 1076 &$-$5 & 0.109 & 1.178~ 0.010 & 18.26~ 0.27 & 1.65 & 0.992 & 18.648 & 1.026 & 18.775 &  1.990~ 0.039 & 0.251~ 0.007 \\
N4291    & 1715 &$-$5 & 0.158 & 1.175~ 0.017 & 20.35~ 0.37 & 1.68 & 1.294 & 19.154 & 1.378 & 19.489 &  2.406~ 0.055 & 0.299~ 0.010 \\
N4339    & 1298 &$-$5 & 0.110 & 1.200~ 0.015 & 19.35~ 0.26 & 2.07 & 1.439 & 19.994 & 1.575 & 20.488 &  1.964~ 0.039 & 0.242~ 0.007 \\
N4346    &  762 &$-$2 & 0.056 & 1.158~ 0.012 & 19.00~ 0.25 & 1.00 & 1.254 & 18.918 & 1.411 & 19.433 &  \dots~ \dots & \dots~ \dots \\
N4365    & 1240 &$-$5 & 0.092 & 1.222~ 0.017 & 21.66~ 0.24 & 1.35 & 1.735 & 19.755 & 1.771 & 19.849 &  2.393~ 0.009 & 0.307~ 0.003 \\
N4373    & 3392 &$-$3 & 0.348 & 1.135~ 0.023 & 21.96~ 0.25 & 2.81 & 1.533 & 19.651 & 1.557 & 19.752 &  2.384~ 0.019 & 0.286~ 0.008 \\
N4374    & 1033 &$-$5 & 0.175 & 1.191~ 0.008 & 21.81~ 0.22 & 1.30 & 1.776 & 19.427 & 1.721 & 19.201 &  2.448~ 0.006 & 0.290~ 0.002 \\
N4379    & 1071 &$-$3 & 0.101 & 1.185~ 0.017 & 18.70~ 0.44 & 1.42 & 1.196 & 19.043 & 1.209 & 19.102 &  \dots~ \dots & \dots~ \dots \\
N4382    &  758 &$-$1 & 0.134 & 1.150~ 0.022 & 22.09~ 0.22 & 0.76 & 2.029 & 20.216 & 2.179 & 20.674 &  2.207~ 0.015 & 0.193~ 0.006 \\
N4386    & 1649 &$-$2 & 0.166 & 1.196~ 0.019 & 20.30~ 0.52 & 1.87 & 1.310 & 19.455 & 1.440 & 19.932 &  \dots~ \dots & \dots~ \dots \\
N4387    &  583 &$-$5 & 0.143 & 1.163~ 0.011 & 18.92~ 0.75 & 1.46 & 1.069 & 18.971 & 1.030 & 18.842 &  1.918~ 0.031 & 0.208~ 0.005 \\
N4391    & 1337 &$-$3 & 0.078 & 1.112~ 0.015 & 18.27~ 0.33 & 1.38 & 1.076 & 19.633 & 1.302 & 20.416 &  \dots~ \dots & \dots~ \dots \\
N4406    &$-$221 &$-$5 & 0.128 & 1.167~ 0.008 & 21.82~ 0.23 & 0.85 & 1.991 & 20.235 & 2.066 & 20.491 &  2.318~ 0.008 & 0.247~ 0.003 \\
N4419    &$-$182 &  1 & 0.143 & 1.126~ 0.026 & 19.03~ 0.29 & 1.15 & 1.296 & 18.816 & 1.293 & 18.966 &  \dots~ \dots & \dots~ \dots \\
N4434    & 1068 &$-$5 & 0.096 & 1.125~ 0.015 & 19.20~ 0.25 & 0.61 & 1.075 & 19.017 & 1.122 & 19.128 &  2.079~ 0.016 & 0.252~ 0.004 \\
N4441    & 1439 &$-$1 & 0.086 & 1.005~ 0.012 & 17.23~ 0.48 & 1.51 & 1.206 & 20.286 & 1.199 & 20.306 &  \dots~ \dots & \dots~ \dots \\
N4458    &  662 &$-$5 & 0.103 & 1.140~ 0.011 & 18.36~ 0.22 & 0.66 & 1.222 & 19.709 & 1.313 & 19.983 &  1.975~ 0.016 & 0.203~ 0.004 \\
N4460    &  558 &$-$1 & 0.082 & 1.011~ 0.015 & 16.80~ 0.26 & 0.88 & 1.393 & 20.204 & 1.861 & 21.744 &  \dots~ \dots & \dots~ \dots \\
N4468    &  895 &$-$3 & 0.198 & 1.045~ 0.015 & 17.40~ 0.23 & 1.35 & 1.495 & 21.378 & 1.288 & 20.674 &  \dots~ \dots & 0.141~ 0.010 \\
N4472    &  997 &$-$5 & 0.096 & 1.218~ 0.011 & 22.18~ 0.21 & 1.08 & 1.796 & 19.020 & 1.846 & 19.187 &  2.452~ 0.006 & 0.299~ 0.002 \\
N4473    & 2236 &$-$5 & 0.123 & 1.158~ 0.012 & 20.37~ 0.23 & 1.62 & 1.542 & 19.169 & 1.569 & 19.284 &  2.255~ 0.009 & 0.299~ 0.003 \\
N4476    & 1955 &$-$3 & 0.123 & 1.048~ 0.017 & 17.89~ 0.24 & 1.40 & 1.234 & 19.744 & 1.210 & 19.700 &  1.564~ 0.039 & 0.148~ 0.007 \\
N4478    & 1370 &$-$5 & 0.107 & 1.164~ 0.019 & 19.35~ 0.33 & 1.49 & 1.097 & 18.320 & 1.254 & 18.896 &  2.122~ 0.007 & 0.245~ 0.003 \\
N4486    & 1292 &$-$4 & 0.096 & 1.244~ 0.012 & 22.11~ 0.24 & 0.74 & 1.793 & 19.192 & 1.779 & 19.178 &  2.535~ 0.009 & 0.282~ 0.004 \\
N4489    &  960 &$-$5 & 0.121 & 1.046~ 0.015 & 17.93~ 0.24 & 1.48 & 1.230 & 19.776 & 1.408 & 20.401 &  1.692~ 0.010 & 0.191~ 0.004 \\
N4494    & 1350 &$-$5 & 0.093 & 1.139~ 0.010 & 20.69~ 0.22 & 0.91 & 1.638 & 19.435 & 1.724 & 19.738 &  2.086~ 0.055 & 0.262~ 0.010 \\
N4526    &  602 &$-$2 & 0.096 & 1.188~ 0.021 & 21.06~ 0.26 & 0.92 & 1.587 & 19.033 & 1.511 & 18.762 &  \dots~ \dots & \dots~ \dots \\
N4546    & 1037 &$-$3 & 0.147 & 1.155~ 0.013 & 19.67~ 0.28 & 1.20 & 1.403 & 18.945 & 1.403 & 18.992 &  \dots~ \dots & \dots~ \dots \\
N4550    &  381 &$-$2 & 0.168 & 1.078~ 0.011 & 18.41~ 0.27 & 1.11 & 1.118 & 18.659 & 1.149 & 18.833 &  \dots~ \dots & \dots~ \dots \\
N4551    & 1198 &$-$5 & 0.166 & 1.170~ 0.009 & 18.81~ 0.26 & 1.05 & 1.171 & 19.151 & 1.150 & 19.078 &  2.011~ 0.016 & 0.251~ 0.004 \\
N4552    &  322 &$-$5 & 0.177 & 1.194~ 0.015 & 20.49~ 0.23 & 1.35 & 1.437 & 18.682 & 1.461 & 18.674 &  2.382~ 0.008 & 0.297~ 0.003 \\
N4564    & 1165 &$-$5 & 0.151 & 1.161~ 0.009 & 19.28~ 0.26 & 1.12 & 1.264 & 18.796 & 1.337 & 19.030 &  2.201~ 0.009 & 0.310~ 0.003 \\
N4565    & 1227 &  3 & 0.067 & 1.128~ 0.027 & 21.04~ 0.23 & 0.99 & 1.834 & 20.050 & 1.873 & 20.457 &  \dots~ \dots & \dots~ \dots \\
N4578    & 2284 &$-$2 & 0.092 & 1.127~ 0.015 & 19.66~ 0.22 & 1.12 & 1.706 & 20.894 & 1.840 & 21.357 &  \dots~ \dots & \dots~ \dots \\
\noalign{\vspace{3pt}\hrule\vspace{3pt}}
\end{tabular}\end{minipage}\end{table*}
\newpage
\begin{table*}\setcounter{table}{0}
\centering \begin{minipage}{172mm}
\caption{Fundamental Plane Data for SBF Survey Galaxies\,---\,\rlap{\it continued}}
\tabcolsep=0.15cm\footnotesize
\begin{tabular}{lrrcccccccccc}
\noalign{\vspace{3pt}\hrule\vspace{3pt}}
Galaxy & $v_h\;$ & $T$ & $A_B$ & \viz~ $\pm$~ & ~\Nbar~~~ $\pm$ & \PD\ &
 $\log R_e^{(I)}$ & $\langle \mu_R \rangle_e^{(I)}$ & $\log R_e^{(V)}$ & $\langle \mu_R \rangle_e^{(V)}$ &
 $\log\sigma$~~ $\pm$~~ & Mg$_2$~~~ $\pm$~ \\
\noalign{\vspace{3pt}\hrule\vspace{3pt}}
N4594    & 1128 &  1 & 0.221 & 1.175~ 0.031 & \dots~ \dots& 1.04 & \dots &  \dots & 1.994 & 19.044 &  2.361~ 0.022 & 0.310~ 0.009 \\
N4600    &  787 &$-$2 & 0.116 & 1.141~ 0.017 & 16.08~ 0.28 & 1.66 & 1.210 & 20.075 & 1.205 & 20.032 &  \dots~ \dots & \dots~ \dots \\
N4616    & 4585 &$-$4 & 0.549 & 1.196~ 0.021 & 20.31~ 0.29 & 2.36 & 1.408 & 20.701 & 1.184 & 20.111 &  2.243~ 0.031 & 0.267~ 0.014 \\
N4620    & 1178 &$-$2 & 0.126 & 1.048~ 0.019 & 17.89~ 0.35 & 1.51 & 1.381 & 20.951 & 2.091 & 23.017 &  \dots~ \dots & \dots~ \dots \\
N4621    &  444 &$-$5 & 0.144 & 1.172~ 0.018 & 21.22~ 0.27 & 1.62 & 1.697 & 19.520 & 1.672 & 19.464 &  2.336~ 0.014 & 0.308~ 0.004 \\
N4636    &  937 &$-$5 & 0.120 & 1.233~ 0.012 & 21.45~ 0.23 & 1.29 & 1.974 & 20.493 & 2.185 & 21.091 &  2.297~ 0.009 & 0.300~ 0.003 \\
N4638    & 1148 &$-$3 & 0.110 & 1.149~ 0.013 & 19.89~ 0.32 & 1.23 & 1.190 & 18.542 & 1.246 & 18.758 &  \dots~ \dots & \dots~ \dots \\
N4645    & 2593 &$-$4 & 0.639 & 1.188~ 0.025 & 20.66~ 0.24 & 2.13 & 1.364 & 19.531 & 1.371 & 19.557 &  2.283~ 0.009 & 0.278~ 0.004 \\
N4649    & 1095 &$-$5 & 0.115 & 1.232~ 0.023 & 21.83~ 0.22 & 0.86 & 1.619 & 18.634 & 1.630 & 18.702 &  2.531~ 0.008 & 0.329~ 0.003 \\
N4684    & 1589 &$-$1 & 0.119 & 1.100~ 0.015 & 18.35~ 0.26 & 2.68 & 1.259 & 19.152 & 1.203 & 19.030 &  \dots~ \dots & \dots~ \dots \\
N4696    & 2958 &$-$4 & 0.490 & 1.203~ 0.014 & 22.40~ 0.25 & 2.53 & 1.760 & 20.212 & 1.774 & 20.277 &  2.398~ 0.018 & 0.276~ 0.007 \\
N4697    & 1210 &$-$5 & 0.132 & 1.157~ 0.010 & 20.71~ 0.23 & 2.19 & 1.850 & 19.749 & 1.822 & 19.650 &  2.215~ 0.008 & 0.278~ 0.004 \\
N4709    & 4650 &$-$5 & 0.513 & 1.199~ 0.011 & 21.77~ 0.30 & 1.80 & 1.776 & 20.860 & 1.896 & 21.236 &  2.388~ 0.008 & 0.328~ 0.003 \\
N4725    & 1207 &  2 & 0.051 & 1.209~ 0.023 & 20.71~ 0.38 & 1.82 & 1.898 & 20.516 & 1.933 & 20.677 &  \dots~ \dots & \dots~ \dots \\
N4729    & 3293 &$-$4 & 0.543 & 1.219~ 0.028 & 20.34~ 0.34 & 3.04 & 1.454 & 20.451 & 1.428 & 20.367 &  2.154~ 0.009 & 0.276~ 0.004 \\
N4742    & 1321 &$-$5 & 0.176 & 1.045~ 0.017 & 18.47~ 0.24 & 1.84 & 1.164 & 18.588 & 1.158 & 18.421 &  1.969~ 0.024 & 0.158~ 0.006 \\
N4754    & 1374 &$-$3 & 0.143 & 1.178~ 0.011 & 20.27~ 0.23 & 1.82 & 1.539 & 19.510 & 1.661 & 19.921 &  \dots~ \dots & \dots~ \dots \\
N4767    & 3005 &$-$5 & 0.461 & 1.175~ 0.028 & 20.85~ 0.29 & 2.80 & 1.329 & 19.289 & 1.361 & 19.377 &  2.319~ 0.016 & 0.291~ 0.005 \\
N4802    & 1013 &$-$2 & 0.206 & 1.004~ 0.018 & 17.66~ 0.24 & 1.63 & 1.347 & 19.477 & 1.344 & 19.415 &  \dots~ \dots & \dots~ \dots \\
N4826    &  414 &  2 & 0.178 & 1.029~ 0.011 & 19.88~ 0.27 & 0.77 & 1.816 & 18.803 & \dots &  \dots &  \dots~ \dots & \dots~ \dots \\
N4946    & 3036 &$-$4 & 0.475 & 1.190~ 0.021 & 21.33~ 0.24 & 2.45 & 1.449 & 20.295 & 1.359 & 19.964 &  2.309~ 0.025 & 0.303~ 0.010 \\
N5011    & 3099 &$-$5 & 0.427 & 1.189~ 0.027 & 21.95~ 0.27 & 2.73 & 1.595 & 20.112 & 1.466 & 19.645 &  2.390~ 0.017 & 0.272~ 0.006 \\
N5044    & 2704 &$-$5 & 0.301 & 1.210~ 0.027 & 21.23~ 0.32 & 2.18 & 1.406 & 19.367 & 1.412 & 19.422 &  2.365~ 0.039 & 0.320~ 0.007 \\
N5061    & 2040 &$-$5 & 0.296 & 1.099~ 0.023 & 21.07~ 0.30 & 3.13 & 1.667 & 19.666 & \dots &  \dots &  2.301~ 0.022 & 0.241~ 0.006 \\
N5090    & 3376 &$-$5 & 0.621 & 1.210~ 0.032 & 22.64~ 0.39 & 2.77 & 2.027 & 21.463 & 2.013 & 21.360 &  2.440~ 0.021 & 0.315~ 0.007 \\
N5193    & 3644 &$-$5 & 0.243 & 1.164~ 0.034 & 20.54~ 0.32 & 2.91 & 1.371 & 19.832 & 1.323 & 19.677 &  2.338~ 0.019 & 0.300~ 0.008 \\
N5273    & 1089 &$-$2 & 0.043 & 1.142~ 0.017 & 19.14~ 0.31 & 1.38 & 1.593 & 20.694 & 2.049 & 22.049 &  \dots~ \dots & \dots~ \dots \\
N5322    & 1804 &$-$5 & 0.061 & 1.183~ 0.011 & 22.02~ 0.30 & 1.74 & 1.723 & 20.042 & 1.490 & 19.168 &  2.342~ 0.039 & 0.265~ 0.007 \\
N5338    &  777 &$-$2 & 0.118 & 1.019~ 0.023 & 16.57~ 0.34 & 2.05 & 1.477 & 21.524 & 1.496 & 21.565 &  \dots~ \dots & \dots~ \dots \\
N5485    & 1985 &$-$2 & 0.072 & 1.180~ 0.017 & 20.35~ 0.39 & 2.32 & 1.440 & 19.871 & 1.419 & 19.825 &  2.185~ 0.055 & 0.283~ 0.010 \\
N5557    & 3258 &$-$5 & 0.025 & 1.202~ 0.021 & 21.30~ 0.34 & 3.48 & 1.488 & 19.768 & 1.401 & 19.455 &  2.412~ 0.039 & 0.305~ 0.007 \\
N5574    & 1582 &$-$3 & 0.135 & 1.054~ 0.011 & 18.34~ 0.64 & 2.18 & 1.036 & 19.053 & 1.096 & 19.235 &  \dots~ \dots & \dots~ \dots \\
N5576    & 1555 &$-$5 & 0.136 & 1.098~ 0.010 & 20.60~ 0.23 & 1.71 & 1.585 & 19.909 & 1.545 & 19.722 &  2.264~ 0.055 & 0.240~ 0.010 \\
N5582    & 1314 &$-$5 & 0.059 & 1.145~ 0.013 & 20.12~ 0.30 & 1.21 & 1.526 & 20.573 & 1.537 & 20.635 &  2.124~ 0.023 & 0.274~ 0.006 \\
N5631    & 1950 &$-$2 & 0.087 & 1.122~ 0.019 & 20.27~ 0.30 & 2.41 & 1.372 & 19.467 & 1.479 & 19.865 &  \dots~ \dots & \dots~ \dots \\
N5638    & 1648 &$-$5 & 0.143 & 1.169~ 0.011 & 20.43~ 0.30 & 1.79 & 1.407 & 19.605 & 1.381 & 19.506 &  2.192~ 0.009 & 0.301~ 0.004 \\
N5687    & 2200 &$-$3 & 0.048 & 1.174~ 0.013 & 20.26~ 0.54 & 2.68 & 1.614 & 20.965 & 1.704 & 21.296 &  \dots~ \dots & \dots~ \dots \\
N5770    & 1464 &$-$2 & 0.167 & 1.120~ 0.017 & 18.48~ 0.34 & 1.97 & 1.160 & 19.375 & \dots &  \dots &  \dots~ \dots & \dots~ \dots \\
N5812    & 2066 &$-$5 & 0.376 & 1.213~ 0.015 & 21.18~ 0.34 & 2.35 & 1.608 & 20.138 & 1.596 & 20.010 &  2.301~ 0.009 & 0.313~ 0.004 \\
N5813    & 1963 &$-$5 & 0.246 & 1.189~ 0.014 & 22.06~ 0.26 & 1.20 & 1.847 & 20.709 & 1.980 & 21.177 &  2.356~ 0.009 & 0.297~ 0.004 \\
N5831    & 1683 &$-$5 & 0.257 & 1.140~ 0.010 & 20.32~ 0.25 & 1.14 & 1.478 & 20.002 & 1.505 & 20.099 &  2.211~ 0.009 & 0.278~ 0.004 \\
N5839    & 1211 &$-$2 & 0.227 & 1.190~ 0.011 & 19.12~ 0.36 & 1.24 & 1.215 & 19.749 & 1.432 & 20.532 &  \dots~ \dots & \dots~ \dots \\
N5846    & 1709 &$-$5 & 0.238 & 1.227~ 0.007 & 21.90~ 0.28 & 1.02 & 1.742 & 19.996 & 1.725 & 19.930 &  2.365~ 0.008 & 0.313~ 0.003 \\
N5866    &  672 &$-$1 & 0.058 & 1.121~ 0.009 & 20.27~ 0.22 & 0.87 & 1.566 & 19.156 & 1.563 & 19.230 &  \dots~ \dots & \dots~ \dots \\
N5869    & 2110 &$-$2 & 0.233 & 1.161~ 0.014 & 19.73~ 1.24 & 2.24 & 1.290 & 19.556 & 1.321 & 19.698 &  \dots~ \dots & \dots~ \dots \\
N5898    & 2090 &$-$5 & 0.629 & 1.169~ 0.009 & 21.05~ 0.32 & 2.12 & 1.466 & 19.511 & 1.506 & 19.659 &  2.353~ 0.013 & 0.302~ 0.004 \\
N5903    & 2518 &$-$5 & 0.638 & 1.142~ 0.011 & 21.37~ 0.30 & 2.28 & 1.629 & 20.182 & 1.629 & 20.174 &  2.306~ 0.016 & 0.287~ 0.005 \\
N5982    & 2936 &$-$5 & 0.078 & 1.166~ 0.010 & 21.77~ 0.68 & 2.83 & 1.537 & 19.986 & 1.558 & 20.067 &  2.385~ 0.023 & 0.285~ 0.006 \\
N6305    & 2770 &$-$3 & 0.441 & 1.147~ 0.051 & 18.92~ 0.28 & 2.33 & 1.206 & 19.308 & 1.055 & 18.764 &  2.242~ 0.041 & 0.249~ 0.014 \\
N6411    & 3690 &$-$5 & 0.249 & 1.156~ 0.008 & 20.13~ 0.40 & 3.35 & 1.467 & 20.342 & 1.429 & 20.238 &  2.236~ 0.016 & 0.269~ 0.004 \\
N6483    & 4880 &$-$5 & 0.257 & 1.152~ 0.051 & 20.62~ 0.47 & 3.33 & 1.272 & 19.850 & 1.122 & 19.285 &  2.353~ 0.042 & 0.277~ 0.010 \\
N6502    & 4880 &$-$4 & 0.267 & 1.191~ 0.041 & 20.84~ 0.30 & 3.81 & 1.398 & 20.367 & 1.259 & 19.851 &  \dots~ \dots & \dots~ \dots \\
N6548    & 2169 &$-$2 & 0.353 & 1.244~ 0.009 & 20.68~ 0.28 & 1.77 & 1.655 & 20.686 & 1.918 & 21.531 &  \dots~ \dots & \dots~ \dots \\
N6684    &  865 &$-$2 & 0.291 & 1.116~ 0.015 & 20.05~ 0.30 & 1.07 & 1.719 & 19.912 & 2.002 & 20.814 &  2.006~ 0.013 & 0.222~ 0.004 \\
N6702    & 4725 &$-$5 & 0.473 & 1.149~ 0.006 & 21.07~ 0.30 & 2.71 & 1.401 & 20.193 & 1.527 & 20.634 &  2.250~ 0.008 & 0.254~ 0.004 \\
N6703    & 2370 &$-$3 & 0.383 & 1.164~ 0.006 & 20.85~ 0.35 & 2.02 & 1.590 & 20.097 & 1.598 & 20.123 &  2.266~ 0.008 & 0.267~ 0.003 \\
N6851    & 3043 &$-$5 & 0.202 & 1.137~ 0.016 & 20.49~ 0.28 & 2.04 & 1.253 & 19.286 & 1.235 & 19.223 &  2.267~ 0.042 & 0.279~ 0.010 \\
N6861    & 2819 &$-$3 & 0.234 & 1.221~ 0.019 & 21.04~ 0.40 & 2.47 & 1.359 & 19.149 & 1.312 & 19.005 &  \dots~ \dots & \dots~ \dots \\
N6868    & 2876 &$-$5 & 0.237 & 1.230~ 0.009 & 21.43~ 0.31 & 2.77 & 1.479 & 19.304 & 1.501 & 19.385 &  2.464~ 0.030 & 0.333~ 0.007 \\
N6869    & 2739 &$-$2 & 0.790 & 1.164~ 0.007 & 20.62~ 0.29 & 2.34 & 1.375 & 19.658 & 1.285 & 19.352 &  \dots~ \dots & \dots~ \dots \\
N6876    & 3951 &$-$5 & 0.195 & 1.178~ 0.009 & 21.89~ 0.43 & 3.36 & 1.429 & 19.520 & 1.467 & 19.650 &  2.374~ 0.030 & 0.300~ 0.007 \\
N6909    & 2753 &$-$4 & 0.164 & 1.064~ 0.020 & 19.95~ 0.23 & 2.00 & 1.420 & 20.267 & 1.231 & 19.585 &  2.178~ 0.042 & 0.218~ 0.010 \\
N7029    & 2818 &$-$5 & 0.159 & 1.138~ 0.032 & 20.77~ 0.50 & 2.27 & 1.249 & 19.133 & 1.270 & 19.189 &  2.308~ 0.030 & 0.266~ 0.007 \\
N7041    & 1877 &$-$3 & 0.173 & 1.139~ 0.055 & 20.21~ 0.27 & 1.66 & 1.347 & 19.337 & \dots &  \dots &  \dots~ \dots & \dots~ \dots \\
\noalign{\vspace{3pt}\hrule\vspace{3pt}}
\end{tabular}\end{minipage}\end{table*}
\newpage
\begin{table*}\setcounter{table}{0}
\centering \begin{minipage}{172mm}
\caption{Fundamental Plane Data for SBF Survey Galaxies\,---\,\rlap{\it continued}}
\tabcolsep=0.15cm\footnotesize
\begin{tabular}{lrrcccccccccc}
\noalign{\vspace{3pt}\hrule\vspace{3pt}}
Galaxy & $v_h\;$ & $T$ & $A_B$ & \viz~ $\pm$~ & ~\Nbar~~~ $\pm$ & \PD\ &
 $\log R_e^{(I)}$ & $\langle \mu_R \rangle_e^{(I)}$ & $\log R_e^{(V)}$ & $\langle \mu_R \rangle_e^{(V)}$ &
 $\log\sigma$~~ $\pm$~~ & Mg$_2$~~~ $\pm$~ \\
\noalign{\vspace{3pt}\hrule\vspace{3pt}}
N7049    & 2158 &$-$2 & 0.237 & 1.174~ 0.033 & 21.54~ 0.25 & 1.58 & 1.448 & 19.002 & 1.483 & 19.237 &  \dots~ \dots & \dots~ \dots \\
N7097    & 2404 &$-$5 & 0.085 & 1.176~ 0.022 & 20.26~ 0.27 & 1.55 & 1.182 & 19.133 & 1.191 & 19.192 &  \dots~ \dots & \dots~ \dots \\
N7144    & 1921 &$-$5 & 0.091 & 1.161~ 0.009 & 20.63~ 0.22 & 1.65 & 1.592 & 20.140 & 1.652 & 20.361 &  2.272~ 0.042 & 0.291~ 0.010 \\
N7145    & 1874 &$-$5 & 0.091 & 1.133~ 0.016 & 20.12~ 0.28 & 1.76 & 1.583 & 20.372 & 1.544 & 20.236 &  2.125~ 0.030 & 0.258~ 0.007 \\
N7168    & 2747 &$-$5 & 0.101 & 1.184~ 0.022 & 20.50~ 0.29 & 2.36 & 1.336 & 19.870 & 1.310 & 19.756 &  \dots~ \dots & \dots~ \dots \\
N7173    & 2501 &$-$4 & 0.114 & 1.150~ 0.017 & 19.91~ 0.26 & 1.89 & 1.168 & 19.211 & 1.164 & 19.207 &  2.293~ 0.030 & 0.280~ 0.007 \\
N7180    & 1479 &$-$2 & 0.138 & 1.109~ 0.009 & 17.86~ 0.32 & 1.35 & 1.015 & 19.083 & 1.178 & 19.644 &  1.987~ 0.026 & 0.187~ 0.006 \\
N7185    & 1838 &$-$3 & 0.135 & 1.080~ 0.012 & 18.44~ 0.41 & 2.18 & 1.474 & 20.884 & 1.628 & 21.430 &  \dots~ \dots & \dots~ \dots \\
N7192    & 2879 &$-$4 & 0.146 & 1.174~ 0.032 & 21.17~ 0.34 & 2.60 & 1.401 & 19.633 & 1.414 & 19.681 &  2.275~ 0.042 & 0.263~ 0.010 \\
N7196    & 3007 &$-$5 & 0.095 & 1.211~ 0.022 & 21.71~ 0.34 & 1.86 & 1.389 & 19.604 & 1.414 & 19.685 &  2.451~ 0.042 & 0.302~ 0.010 \\
N7200    & 2897 &$-$4 & 0.083 & 1.183~ 0.018 & 19.24~ 0.37 & 2.03 & 1.062 & 19.741 & 1.026 & 19.527 &  2.299~ 0.042 & 0.296~ 0.010 \\
N7280    & 1903 &$-$1 & 0.240 & 1.105~ 0.009 & 19.56~ 0.29 & 1.67 & 1.576 & 20.823 & 1.950 & 21.989 &  \dots~ \dots & \dots~ \dots \\
N7302    & 2586 &$-$3 & 0.299 & 1.126~ 0.016 & 19.19~ 0.28 & 2.33 & 1.201 & 19.368 & 1.578 & 20.692 &  \dots~ \dots & \dots~ \dots \\
N7331    &  819 &  3 & 0.392 & 1.120~ 0.017 & 20.75~ 0.24 & 0.36 & 1.702 & 19.035 & 1.805 & 19.521 &  \dots~ \dots & \dots~ \dots \\
N7332    & 1207 &$-$2 & 0.161 & 1.107~ 0.008 & 19.98~ 0.28 & 0.61 & 1.242 & 18.648 & 1.252 & 18.639 &  \dots~ \dots & \dots~ \dots \\
N7454    & 2007 &$-$5 & 0.343 & 1.123~ 0.012 & 19.66~ 0.35 & 1.23 & 1.474 & 20.278 & 1.325 & 19.753 &  2.018~ 0.009 & 0.199~ 0.004 \\
N7457    &  822 &$-$3 & 0.228 & 1.104~ 0.009 & 18.99~ 0.28 & 0.36 & 1.626 & 20.343 & 1.831 & 21.015 &  \dots~ \dots & \dots~ \dots \\
N7507    & 1548 &$-$5 & 0.206 & 1.196~ 0.009 & 21.54~ 0.25 & 1.20 & 1.600 & 19.494 & 1.472 & 19.000 &  2.369~ 0.033 & 0.324~ 0.007 \\
N7562    & 3608 &$-$5 & 0.453 & 1.187~ 0.018 & 21.95~ 0.85 & 2.97 & 1.295 & 19.278 & 1.386 & 19.486 &  2.394~ 0.009 & 0.280~ 0.004 \\
N7619    & 3747 &$-$5 & 0.347 & 1.229~ 0.009 & 22.74~ 0.37 & 2.46 & 1.656 & 20.330 & 1.680 & 20.414 &  2.510~ 0.009 & 0.333~ 0.003 \\
N7743    & 1658 &$-$1 & 0.305 & 1.080~ 0.009 & 19.51~ 0.25 & 1.03 & 1.522 & 20.145 & 2.105 & 22.072 &  \dots~ \dots & \dots~ \dots \\
N7796    & 3252 &$-$4 & 0.044 & 1.230~ 0.022 & 22.21~ 0.43 & 2.43 & 1.571 & 20.325 & 1.608 & 20.450 &  \dots~ \dots & \dots~ \dots \\
N7814    & 1050 &  2 & 0.192 & 1.245~ 0.017 & 20.23~ 0.22 & 0.51 & 1.611 & 19.734 & 1.571 & 19.740 &  \dots~ \dots & \dots~ \dots \\
U07767   & 1331 &$-$5 & 0.089 & 1.152~ 0.019 & \dots~ \dots& 1.67 & \dots &  \dots & 1.029 & 19.611 &  \dots~ \dots & \dots~ \dots \\
\noalign{\vspace{3pt}\hrule\vspace{3pt}}
\end{tabular}\end{minipage}\end{table*}

\bigskip

\begin{table*}
\begin{minipage}{120mm}
\caption{\xfp\ Comparisons \label{tab:xfpres}}
\tabcolsep=0.35cm\small
\begin{tabular}{ccrccc}
\hline
{Comparison} & {Galaxy}  & $N_g$ & {$\langle \Delta \xfp\rangle $} & 
rms\footnote[1]{Dispersion for comparisons of $\xfp\equiv\log R_e - 0.33\sbe\,$.} & 
rms(0.30)\footnote[2]{Dispersion for comparisons of $\xfp^{\prime}\equiv\log R_e - 0.30\sbe\,$.}\\
 & (type) &   & (dex) & (dex) & (dex) \\
\hline
VSBF $-$ ISBF\dotfill & all      & 257 & $+0.002 \pm 0.001$ & 0.023 & 0.019 \\
VSBF $-$ ISBF, no N4111\dotfill&all& 256 & $+0.002 \pm 0.001$ & 0.021 & 0.017 \\
VSBF $-$ ISBF\dotfill &   E      & 122 & $-0.001 \pm 0.001$ & 0.016 & 0.012 \\[5pt]
ISBF $-$ SMAC\dotfill & all & 32 & $-0.030 \pm 0.008$ & 0.044 & 0.030 \\
ISBF $-$ SMAC, no N1553\dotfill &  all & 31 & $-0.033 \pm 0.007$ & 0.040 & 0.028\\
ISBF $-$ SMAC\dotfill &   E & 18 & $-0.025 \pm 0.007$ & 0.029 & 0.019 \\[5pt]
VSBF $-$ SMAC\dotfill & all & 32 & $-0.033 \pm 0.008$ & 0.044 & 0.031 \\
VSBF $-$ SMAC, no N1553\dotfill &  all & 31 & $-0.036 \pm 0.007$ & 0.039 & 0.028\\
VSBF $-$ SMAC\dotfill &   E & 18 & $-0.028 \pm 0.007$ & 0.027 & 0.018 \\[5pt]
ISBF $-$ 7S\dotfill & all & 169 & $-0.036\pm0.003$ & 0.042 & 0.031 \\
ISBF $-$ 7S\dotfill &   E & 120 & $-0.033\pm0.004$ & 0.040 & 0.029  \\
ISBF $-$ 7S, red\footnote[3]{`Red' subsample: restricted to galaxies with $\viz\ge1.135\,$.}\dotfill
 & all & 134 & $-0.029\pm0.003$ & 0.033 & 0.025 \\
ISBF $-$ 7S, red{$^c$}\dotfill &   E & 100 & $-0.026\pm0.003$ & 0.028 & 0.020 \\[5pt]
VSBF $-$ 7S\dotfill & all & 165 & $-0.036\pm0.003$ & 0.039 & 0.030 \\
VSBF $-$ 7S\dotfill &   E & 118 & $-0.031\pm0.003$ & 0.036 & 0.027  \\
VSBF $-$ 7S, red{$^c$}\dotfill & all & 132 & $-0.030\pm0.003$ & 0.031 & 0.025 \\
VSBF $-$ 7S, red{$^c$}\dotfill &   E &  99 & $-0.027\pm0.003$ & 0.027 & 0.020 \\[5pt]
7S $-$ SMAC\footnote[4]{7S vs.\ SMAC comparisons restricted to $c{z}<5000$ \kms\
because of problems arising from seeing effects on the 7S data at larger distances.}\dotfill
 & all & 46 & $-0.003\pm0.005$ & 0.036 & 0.030 \\
7S $-$ SMAC{$^d$}\dotfill & E & 25 & $-0.013\pm0.007$ & 0.035 & 0.031 \\
\hline
\end{tabular}\vspace{-12pt}
\end{minipage}\end{table*}

\end{document}